\documentclass[12pt]{article}

\usepackage{amsfonts}
\usepackage{amssymb}
\usepackage{amsthm}
\usepackage[fleqn]{amsmath}
\usepackage[mathscr]{eucal}

\input{epsf}

\makeatletter
\renewcommand \thesection {\@arabic\c@section}
\renewcommand\thesubsection   {\thesection.\@arabic\c@subsection}
\renewcommand\thesubsubsection{\thesubsection .\@arabic\c@subsubsection}
\renewcommand\theparagraph    {\thesubsubsection.\@arabic\c@paragraph}
\renewcommand\section{\@startsection {section}{1}{\z@}%
                                   {-3.5ex \@plus -1ex \@minus -.2ex}%
                                   {1.9ex \@plus.2ex}%
                                   {\normalfont\large\bfseries\centering}}
\renewcommand\subsection{\@startsection{subsection}{2}{\z@}%
                                     {-2ex\@plus -1ex \@minus -.2ex}%
                                     {1.2ex \@plus .2ex}%
                                    {\normalfont\normalsize\bfseries\centering}
}
\renewcommand\subsubsection{\@startsection{subsubsection}{3}{\z@}%
                                     {-2ex\@plus -1ex \@minus -.2ex}%
                                     {.5ex \@plus .2ex}%
                                     {\normalfont\normalsize\em}}
\renewcommand\paragraph{\@startsection{paragraph}{4}{\z@}%
                                    {3.25ex \@plus1ex \@minus.2ex}%
                                    {-1em}%
                                    {\normalfont\normalsize\em}}
\renewcommand\subparagraph{\@startsection{subparagraph}{5}{\parindent}%
                                       {3.25ex \@plus1ex \@minus .2ex}%
                                       {-1em}%
                                      {\normalfont\normalsize\em}}
\makeatother

\newcommand\abs[1]{{\left| #1 \right|}}
\newcommand\norm[1]{{\left\| #1 \right\|}}

\newcounter{subequation}
	\newenvironment{subequation}%
	{\addtocounter{equation}{-1}%
	\stepcounter{subequation}%
	\begin{equation}}%
	{\end{equation}%
}

\newcommand{\beq}{\begin{equation}}
\newcommand{\eeq}{\end{equation}}
\newcommand{\bseq}{\begin{subequation}}
\newcommand{\eseq}{\end{subequation}}
\newcommand{\bea}{\begin{eqnarray}}
\newcommand{\eea}{\end{eqnarray}}

\newcommand{\refeq}[1]{(\ref{#1})}

\newcommand{\supp}{{\mathrm{supp}\, }}
\newcommand{\id}{{\mathrm{id}\,}}

\newcommand{\eps}{\epsilon}
\newcommand{\veps}{\varepsilon}

\newcommand{\cC}{{\cal C}}
\newcommand{\cE}{{\cal E}}
\newcommand{\wcE}{\widetilde{\cal E}}
\newcommand{\cG}{{\cal G}}

\newcommand{\cJ}{{\cal J}}
\newcommand{\cK}{{\cal K}}
\newcommand{\wcK}{\widetilde{\cal K}}

\newcommand{\cP}{{\cal P}}
\newcommand{\cS}{{\cal S}}
\newcommand{\cV}{{\cal V}}

\newcommand{\Rset}{{\mathbb R}}

\newcommand{\Csp}{\mathfrak{C}}
\newcommand{\Hsp}{\mathfrak{H}}

\newcommand{\Lsp}{\mathfrak{L}}

\newcommand{\Psp}{\mathfrak{P}}

\newcommand{\pr}{\prime}

\newcommand{\QED}{$\quad$\textrm{Q.E.D.}\smallskip}

\newcommand{\fe}{\varrho_\veps}

\newcommand{\dd}{{\mathrm{d}}}

\newcommand{\ddt}{\frac{\dd}{\dd{t}}}

\newcommand{\pdt}{{\partial_t^{\phantom{0}}}}
\newcommand{\pdp}{{\nabla_p^{\phantom{0}}}}
\newcommand{\pdq}{{\nabla_q^{\phantom{0}}}}

\newcommand{\pdqsq}{{\Delta_q}}

\newcommand{\rVPa}{{rVP$^-$}}
\newcommand{\rVPr}{{rVP$^+$}}

\newtheorem{defn}{Definition}[section] 

\newtheorem{Coro}[defn]{Corollary}
\newtheorem{Defi}[defn]{Definition}
\newtheorem{Lemm}[defn]{Lemma}
\newtheorem{Prop}[defn]{Proposition}
\newtheorem{Rema}[defn]{Remark}
\newtheorem{Theo}[defn]{Theorem}

\begin{document}

\title{ON THE RELATIVISTIC VLASOV--POISSON SYSTEM}

\author{M.K.-H. KIESSLING and A.S. TAHVILDAR-ZADEH\\
\textit{Department of Mathematics,}\\
\textit{Rutgers, The State University of New Jersey,}\\
\textit{110 Frelinghuysen Rd., Piscataway, NJ 08854}}

\date{}

\maketitle

\begin{abstract}
\noindent
	The Cauchy problem is revisited for the so-called relativistic
Vlasov--Poisson system in the attractive case, originally studied by
Glassey and Schaeffer in 1985.
	It is proved that a unique global classical solution exists 
whenever the positive, integrable initial datum $f_0$ is spherically symmetric,
compactly supported in momentum space, vanishes on characteristics with vanishing 
angular momentum, and its $\Lsp^\beta$ norm is below a critical constant $C_\beta >0$
whenever $\beta\geq {3/2}$.
	It is also shown that, if the bound $C_\beta$ on the $\Lsp^\beta$ norm of 
$f_0$ is replaced by a bound $C>C_\beta$, any $\beta\in (1,\infty)$, then classical 
initial data exist which lead to a blow-up in finite time.
        The sharp value of $C_\beta$ is computed for all $\beta\in (1,3/2]$,
with the results  $C_{\beta}=0$ for $\beta\in(1,3/2)$ and 
$C_{3/2} =\frac{3}{8}\left(\frac{15}{16}\right)^{1/3}$ (when $\norm{f_0}_{\Lsp^1}=1$),
while  for all $\beta > 3/2$ upper and lower bounds on $C_\beta$ are given which 
coincide as $\beta\downarrow 3/2$.
	Thus, the $\Lsp^{3/2}$ bound is optimal in the sense that 
it cannot be weakened to an $\Lsp^\beta$ bound with $\beta < 3/2$, 
whatever that bound.
	A new, non-gravitational physical vindication of the model which 
(unlike the gravitational one) is not restricted to weak fields, is also given.
\end{abstract}

\vfill
\hrule
\smallskip
\noindent
Version of July 08, 2008. Last typos corrected: Jan. 26, 2009. \\
\copyright{2009} The copyright for this preprint resides with the authors. 
Its reproduction, in its entirety, for non-commercial purposes is permitted.
The copyright for the version published in 
\textit{Indiana Univ. Math. J.} \textbf{57}, 3177-3207 (2009) resides with Indiana University 
Mathematics Journal.
\newpage
\section{Introduction}
	In \cite{GlasseySchaefferA}, Glassey and Schaeffer inaugurated a series of
studies (see \cite{GlasseySchaefferC} and the references therein; see also \cite{HadzicRein})
of what they sanctioned the ``relativistic Vlasov--Poisson system'' (rVP in the following).
	The rVP poses a classical Cauchy problem for a relative density function 
$f_t:\Rset^3\times \Rset^3\to\Rset_+$ of an $N$-body system with Cauchy data 
$f_0\in (\Psp\cap\Csp^1)(\dd{p}\dd{q})$,\footnote{By
		$\Psp(\dd{p}\dd{q})$ we denote the probability measures, 
		by $\Psp_n(\dd{p}\dd{q})$ those having finite $n$-th moment, 
		and by $(\Psp_n\cap\Lsp^\alpha)(\dd{p}\dd{q}),\ \alpha\geq 1$, 
		respectively $(\Psp_n\cap\Csp^1)(\dd{p}\dd{q})$, 
		those of these measures which are absolutely 
		continuous w.r.t. Lebesgue measure $\dd{p}\dd{q}$ on $\Rset^3\times\Rset^3$ 
		(momentum$\times$physical space) with density in $\Lsp^\alpha(\dd{p}\dd{q})$, 
                respectively in $\Csp^1(\dd{p}\dd{q})$ which are the functions with one continuous 
                classical derivative
                (abusing notation, we identify these measures with their densities).
		While the relative density function $f_t$ thus fulfills the requirements 
		of a probability density function, it should really be thought of as a
		continuum approximation to a \emph{merely normalized} (i.e. relative)
		empirical ``density'' on $(p,q)$-space of an actual individual $N$-body system. 
                Incidentally, the conspicuous absence of $N$ in	\refeq{rVPfEQ}--\refeq{phiASqTOoo} 
		means that we study the evolution on suitable time and space scales.	
                The scaling transformation 
		$t\mapsto N^{-1}t$,\ $q\mapsto N^{-1}q$, so that $v\mapsto v$ and $p\mapsto p$,
		together with $f\mapsto N^{3} f$ and $\phi\mapsto \phi$ restores $N$ explicitly in 
		\refeq{rVPfEQ},\refeq{rVPphiEQ},\refeq{phiASqTOoo}. 
                Note that $(\Psp\cap\Lsp^1)(\dd{p}\dd{q})$ is invariant under this scaling map.}
given in form of the kinetic equation 
\beq
 \Big(\pdt + v \cdot\pdq +\sigma\pdq\phi_t(q)\cdot\pdp\Big)f_t(p,q) = 0,
\label{rVPfEQ}
\eeq
in which the velocity $v\in\Rset^3$ and momentum $p\in\Rset^3$ of a (point) particle of unit mass
are related by Einstein's formula (with the speed of light $c=1$),
\beq
 v = \frac{p}{\sqrt{1+|p|^2}}\,;
\label{EINSTEINvOFp}
\eeq
the scalar field $\phi_t:\Rset^3\to\Rset_-$ satisfies the Poisson equation
\beq
 \pdqsq\phi_t(q) = 4\pi \int_{\Rset^3}\! f_t(p,q)\dd{p}
\label{rVPphiEQ}
\eeq
with asymptotic condition\footnote{In principle, asymptotic conditions other than 
		\refeq{phiASqTOoo} can be imposed, for instance other harmonic 
		behavior indicating ``system-external sources at infinity.''}
\beq
 \phi_t(q) \asymp -|q|^{-1}
\,
\label{phiASqTOoo}
\eeq
when $|q|\to\infty$, so that\footnote{Note that $\phi_t = - |\id|^{-1}*\int\! f_t \dd{p}$
		does \emph{not} represent dynamical degrees of freedom beyond those of $f_t$.
		Thus, we will speak of solutions $f_t$ of \refeq{rVPfEQ}--\refeq{phiASqTOoo}.}
$\phi_t = - |\id|^{-1}*\int\! f_t \dd{p}$; 
and $\sigma\in\{-1,+1\}$ decides whether the gradient force field of $\phi$ is attractive 
($\sigma=-1$) or repulsive ($\sigma=+1$).
	In this paper we are primarily interested in the attractive case $\sigma=-1$.

	The rVP system is not truly relativistic in the sense of proper Lorentz or even general 
covariance.
	Yet for the mathematically special (and physically idealized) situation of spherical
systems, rVP can actually be obtained from truly relativistic (and physically relevant) Vlasov
models in certain limiting physical regimes; hence, rVP may have some physical significance.
	Its version with $\sigma = +1$ (denoted \rVPr) is obtained directly from the relativistic
Vlasov--Maxwell system (rVM) for a single specie of electrically charged physical particles 
with spherically symmetric initial data without any further conditions; cf. \cite{HorstB}.
	Thus, in the repulsive case $\phi_t(q)$ can be thought of as Coulomb's electrical potential
at the space point $q$ at time $t$.
	The version with $\sigma=-1$ (denoted \rVPa) ought to obtain in a ``weak field limit'' of 
the physically relevant general covariant Vlasov--Einstein system (VE) with spherically symmetric
data, though we are only aware of some work (see \cite{Rendall}) on the combined weak field plus 
low velocity limit which leads to the familiar Vlasov--Poisson system with $\sigma =-1$ (VP$^-$), 
formally obtained from \refeq{rVPfEQ}--\refeq{phiASqTOoo} by replacing Einstein's formula 
\refeq{EINSTEINvOFp} with Newton's $v = p$.
	Thus, in the attractive case $\phi_t(q)$ may be thought of as Newton's
gravitational potential at the space point $q$ at time $t$.
	Unfortunately, this gravitational interpretation has to be taken with a grain of salt, for
some mathematically interesting phenomena such as stationary bound states \cite{Batt, HadzicRein}
and finite-time blow up \cite{GlasseySchaefferA} (signaling gravitational collapse to a 
singularity, see \cite{LMRb}) of this (both psychologically and ``physically'') attractive version of rVP 
occur in the \emph{strong field} regime, i.e. were \rVPa\ can no longer be expected to be a legitimate 
approximation to VE.
	While this would seem to make mathematical studies of the strong field regime
of \rVPa\ questionable from a physical perspective, in the appendix we give an unconventional 
(and presumably surprising) physical interpretation of \rVPa\ with spherical symmetry in terms of 
distributional solutions of rVM for a neutral two-species plasma which is not restricted to 
weak fields, and which could have interesting applications in space physics; thus the mathematically 
rigorous vindication of this electrical interpretation of \rVPa\ is an important open problem.
	
	In the main part of the present paper we revisit the questions of global existence
and uniqueness versus finite time blow-up of solutions to \rVPa, which were addressed already by 
Glassey and Schaeffer \cite{GlasseySchaefferA}.
        We restrict our discussion to classical solutions, but ask for the optimal --- in the sense 
of weakest --- constraints that guarantee that classical data will launch a unique global 
solution of the dynamical system. 
        In this vein, we prove the following result about \rVPa:
\newpage

\begin{Theo}\label{mainTHEOREM}
	A unique global classical solution of \rVPa\ exists for all spherically symmetric 
initial data $f_0\in \Psp_1\cap\Csp^1(\dd{p}\dd{q})$ which are compactly supported 
in momentum space, vanish for $p\times{q}=0$, and satisfy $\norm{f_0}_{3/2} < C_{3/2}$, with 
$C_{3/2} =\frac{3}{8}\left(\frac{15}{16}\right)^{1/3}\approx 0.367$.
	The $\Lsp^{3/2}$ bound $C_{3/2}$ on $f_0$ is optimal in the sense that initial 
data $f_0$ exist which satisfy all the hypotheses except that
$\norm{f_0}_{3/2} > C_{3/2}$, and which lead to a blow-up in finite time.
\end{Theo}
\begin{Rema}
	The critical case $\norm{f_0}_{3/2} = \frac{3}{8}\left(\frac{15}{16}\right)^{1/3}$
is not covered by our theorem.
\end{Rema}
\begin{Rema}
	All global-in-time solutions covered by Theorem \ref{mainTHEOREM} have positive, 
those that blow up in finite time non-positive energy. 
        Among the data that lead to finite time blow-up there are indeed some with zero energy.
	This improves on Glassey--Schaeffer's result that negative energy data will
lead to finite time blow-up.
\end{Rema}
\begin{Rema} 
        By the interpolation inequality, $f_0\in\Psp_1\cap\Lsp^\beta$ with $\beta> 3/2$ 
implies $f_0\in P_1\cap\Lsp^{3/2}$, with $\norm{f_0}_\beta < C_{3/2}^{3(1-1/\beta)}$
implying $\norm{f_0}_{3/2} < C_{3/2}$.
        This shows that a global existence and uniqueness theorem analogous to Theorem 
\ref{mainTHEOREM} can also be stated with the sharp $\Lsp^{3/2}$ condition on $f_0$ replaced 
by a sharp $\Lsp^\beta$ condition on $f_0$ for any $\beta > 3/2$, with $C_{3/2}$ replaced by 
a corresponding sharp constant $C_\beta\geq  C_{3/2}^{3(1-1/\beta)}$. 
        Beside the sharp $C_{3/2}$ given in Theorem \ref{mainTHEOREM}, and the lower bound on
$C_\beta$  for $\beta > 3/2$ just stated, we will also give an explicit upper bound on $C_\beta$ 
for $\beta > 3/2$.
\end{Rema}
\begin{Rema}
        The reverse to the interpolation estimates of course is not true:
$f_0\in\Psp_1\cap\Lsp^{3/2}$ with  $\norm{f_0}_{3/2} < C_{3/2}$ does not imply any bound on 
$\norm{f_0}_\beta$ for $\beta> 3/2$. 
        Thus, our $\Lsp^{3/2}$ condition is weaker than any of the possible $\Lsp^\beta$ conditions 
with $\beta > 3/2$.
        In fact, our $\Lsp^{3/2}$ bound is the weakest possible $\Lsp^\beta$ condition
for which an analog of Theorem \ref{mainTHEOREM} can be formulated, in the sense that
our $\Lsp^{3/2}$ bound on $f_0$ cannot be replaced by an $\Lsp^\beta$ bound with
$\beta \in (1, 3/2)$, whatever that bound.
        Indeed, among the data $f_0$ satisfying any such $\Lsp^\beta$ bound with $\beta<3/2$, there
are some with negative energy, which lead to a blow-up in finite time by Glassey-Schaeffer's blow-up 
theorem (evidently, $\norm{f_0}_{3/2} > C_{3/2}$ for those data).
\end{Rema}
\begin{Rema}
      By the previous two remarks, everything else being equal our sharp $\Lsp^{3/2}$ condition 
on $f_0$ improves on the (nonsharp) $\Lsp^\infty$ condition on $f_0$ in \cite{GlasseySchaefferA}.
\end{Rema}
\begin{Rema}
      If the normalization $\norm{f_0}_1=1$ is changed to any other value for the $\Lsp^1$ norm of $f_0$,
the values of the critical $C_\beta$ for $\beta\geq 3/2$ change by a simple scaling transformation.
\end{Rema}

	The rest of the paper is structured as follows.
	In the next section, we list the familiar conservation laws and the virial identities
for \rVPa. 
	Then, in section 3, we find an $\Lsp^{3/2}$-optimal subset of $\Psp_1\cap\Lsp^1(\dd{p}\dd{q})$ on
which the energy functional of $f$ is bounded below, which bound is 0.
	Section 4 is devoted to obtaining a-priori bounds on the data $f_0$.
	In section 5 we prove our global existence and uniqueness result of classical solutions,
all of which have strictly positive energy.
	In section 6, we prove that blow-up in finite time occurs for certain data with non-positive energy.
        We comment on the critical case in section 7, and 
section 8 lists some interesting open problems.
	Finally, in the appendix we give a vindication of \rVPa\ in terms of certain
distributional solutions to two-species neutral rVM.
\section{Conservation laws and virial identities}
	In our paper we will make use of (many of) the conservation laws, of the virial identity, 
and another identity, all valid a priori for any sufficiently regular solution of \rVPa.
	While (most of) these laws and both identities are proved in \cite{GlasseySchaefferA} (under
more restrictive assumptions than stated here), for the convenience of the reader, these basic results 
are collected separately in this section. 
	To simplify the notation, we will use the abbreviation $\int$ for $\int_{\Rset^3}$, and 
we write
\beq
  \rho(q) := \int f(p,q)\dd{p}
\label{rhoASfINT}
\eeq
for the relative density function in physical space.
	We shall drop the argument $\dd{p}\dd{q}$ from now on from the symbols for the function spaces.

	We begin with the conservation laws for the Casimir functionals of $f$.
	Thus, for (the pertinent subset of) $f\in\Psp\cap \Lsp^1$ we define the 
$g$-Casimir functional of $f$ by
\beq
\cC^{(g)}\left(f\right)
=
{\iint} g\circ f \, \dd{p}\dd{q}\, ,
\quad {\mathrm{for\ all}}\quad g:\Rset_+\to \Rset
\quad {\mathrm{such\ that}}\quad g\circ f\in \Lsp^1
\,.
\label{CASIMIRfuncVeps} 
\eeq 
	For $g(\,\cdot\,) =\, (\id(\,\cdot\,))^\alpha$, $\alpha\geq 1$, we get the $\alpha$-th power of the 
$\Lsp^\alpha$ norm of $f$; when $\alpha=1$ this yields just the mass functional ($=$ integral) of $f$.
	The choice $g(\,\cdot\,) = -\id(\,\cdot\,)  \log(\id(\,\cdot\,) /f_*)$
gives the entropy of $f$ relative to some arbitrary $f_*\in \Psp\cap\Lsp^1$,
\beq
 \cC^{(-{\mathrm{id}} \log({\mathrm{id}}/f_*))}\left(f\right)
=
- \iint f\ln ({f}/{f_*}) \dd{p}\dd{q}
\equiv 
 \cS(f|f_*)\, .
\label{relENTROPY}
\eeq
	Then, since \refeq{rVPfEQ} is isomorphic to a continuity equation for $f$ on $\Rset^6$, 
we have
\begin{Prop}\label{propCASIMIRS} 
	Let $t\mapsto f_t\in\Psp\cap\Csp^1$ be a classical solution of \rVPa.
	Then, whenever $\cC^{(g)}\bigl(f_0\bigr)$ exists, also $\cC^{(g)}\left(f_t\right)$ does, and
\beq
\cC^{(g)}\left(f_t\right) = \cC^{(g)}\bigl(f_0\bigr)
\,.
\label{CASIMIRconst}
\eeq
\end{Prop}
	Beside the conservation laws just stated, the familiar quantities 
energy, momentum, and angular momentum are conserved.
\begin{Prop}\label{propCONSERVATIONlawsA}
	Let $t\mapsto f_t\in\Psp_1\cap\Csp^1$ be a classical solution of \rVPa,
then the energy of $f_t$ is conserved, i.e. $\cE(f_t) =\cE(f_0)$, where
\beq
 \cE(f) 
  :=
   \iint \sqrt{1+ |p|^2} f(p,q) \dd{p}\,\dd{q}
  -
  \frac{1}{2}
 \iiiint\frac{f(p,q)f(p^\prime,q^\prime)}{\abs{q-q^\prime}}\dd{p}\,\dd{q}\,\dd{p^\prime}\,\dd{q^\prime}
\,,
\label{EfuncF}
\eeq
and the momentum space contribution to $\cE(f)$ (denoted $\cE_{p}(f)$)
is the kinetic plus rest energy, while the physical space contribution (denoted $\cE_q(f)$)
is the potential energy of $f$.

	Moreover, the momentum of $f$ is conserved, i.e. $\cP(f_t)=\cP(f_0)$, where
\beq
\cP(f) := \iint p f(p,q)\dd{p}\dd{q}
\,,
\label{linearMOM}
\eeq
and if also $f_t\in\Psp_2$ then so is the angular momentum of $f$,  i.e. $\cJ(f_t)=\cJ(f_0)$,
where
\beq
\cJ(f) := \iint q\times p f(p,q)\dd{p}\dd{q}
\,.
\label{angularMOM}
\eeq
\end{Prop}
	Beside the angular momentum functional $\cJ(f)$, also the virial functional 
\beq
\cV(f):=\iint q\cdot p f(p,q)\dd{p}\dd{q}
\label{virialFUNC}
\eeq
plays an important r\^{o}le, but it is not conserved. 
	Its time evolution yields what is called the dilation identity for \rVPa, which in 
the physics literature would be part of a ``dynamical virial theorem.''
\begin{Prop}\label{propDILATIONid} 
	Let $t\mapsto f_t\in\Psp_2\cap\Csp^1$ be a classical solution of \rVPa\ 
over the interval $(0,T)$.
	Then
\beq
\ddt \cV(f_t)
 =
\cE(f_t)  -  \iint \frac{1}{\sqrt{1+ |p|^2}} f_t(p,q) \dd{p}\,\dd{q}
\label{dilationID}
\eeq
\end{Prop}
	An immediate and entirely obvious corollary of the dilation identity \refeq{dilationID}, 
which nevertheless deserves to be stated in its own right, is the ``stationary virial theorem.''
\begin{Coro}\label{coroVIRIALthm} 
	Let $t\mapsto f_t\equiv f_0$ be a stationary solution of \rVPa.
	Then 
\beq
\cE(f_0) =  \iint \frac{1}{\sqrt{1+ |p|^2}} f_0(p,q) \dd{p}\,\dd{q}
\label{VIRIALidREL}
\,.
\eeq
\end{Coro}
\begin{Rema}
	If the stationary $f_t=f_0$ has most or all of its mass supported in a cylindrical
subset of $(p,q)$ space given by $B_P(0)\times\Rset^3$ with $P\ll 1$, then
we can expand $\sqrt{1+ |p|^2}= 1 +\frac{1}{2}|p|^2 +O(|p|^4)$ in \refeq{VIRIALidREL}, both 
in its r.h.s. and in $\cE(f_0)$, and obtain the familiar  stationary virial theorem
``$2E_{kin}= -E_{pot}$'' of non-relativistic VP$^-$, viz.
\beq
\iint |p|^2 f_0(p,q) \dd{p}\,\dd{q}
=
  \frac{1}{2}
 \iint\frac{\rho(q)\rho(q^\prime)}{\abs{q-q^\prime}}\dd{p^\prime}\,\dd{q^\prime}
\label{VIRIALidNONREL}
\,.
\eeq
\end{Rema}
	Some results will flow entirely from Corollary \ref{coroVIRIALthm}.
	But for the blow-up results we also need 
another part of the virial theorem, identity (26) in \cite{GlasseySchaefferA}, viz.
\begin{Prop}\label{propVIRIALpartII}
	Let $t\mapsto f_t \in\Psp_3\cap\Csp^1$ be a classical solution of \rVPa\ 
over some time interval $t\in (0,T)$.
	Then
\beq
\ddt \iint |q|^2 \sqrt{1+ |p|^2} f_t(p,q) \dd{p}\,\dd{q}
=
2\cV(f_t)
-
\iint  |q|^2 v\cdot\pdq\phi_t(q)f_t(p,q) \dd{p}\,\dd{q}
\,.
\label{virialIDtwo}
\eeq
\end{Prop}
\section{An $\Lsp^\beta$-optimal $f$ domain for lower boundedness of $\cE(f)$}
	The lower boundedness properties of $\cE(f)$ play an important r\^{o}le in
the proof of our theorem. 
	Clearly, $\cE(f)$ is unbounded below on $\Psp_1\cap\Lsp^1$, for we can make $\cE(f)$ 
as negative as we please along the sequence $f_R\in \Psp_1\cap\Lsp^1$, given by
$f_R(p,q) = (16\pi^2/9)^{-1}R^{-3}\chi_{B_1(0)}(p) \times \chi_{B_R(0)}(q)$, by letting
$R\downarrow 0$ (here, $\chi_S$ is the characteristic function of the set $S$).
	So one needs to restrict $\cE(f)$ to some subset of $\Psp_1\cap\Lsp^1$.
	Incidentally, though this is not spelled out in the pertinent references, 
it follows from inequalities (5) and (22) in \cite{GlasseySchaefferA} that $\cE(f)$ is bounded 
below (by 0, then) when $f\in \Psp_1\cap\Lsp^\infty$ with $\norm{f}_\infty$ sufficiently small, 
and it follows from the displayed but unnumbered inequality in the introduction of \cite{HadzicRein} 
that $\cE(f)\geq 0$ also when $f\in \Psp_1\cap \Lsp^\beta$ with $\beta\geq 3/2$ and $\norm{f}_\beta$ 
small enough.
        Neither of these references reveal how small is ``small enough'' and what happens for larger norms 
or smaller $\beta$.

	We here are interested in the $\Lsp^\beta$-optimal domain in $f$ space, in the sense 
that for $f\in \Psp_1\cap\Lsp^\beta$ we seek the smallest possible $\beta$, and the largest 
possible $\Lsp^\beta$ norm of $f$, such that lower boundedness of $\cE(f)$ holds, while 
$\cE(f)$ is unbounded below when these conditions on $f$ are not met.
	If the Laplacian in Poisson's equation \refeq{rVPphiEQ} for $\phi$ is replaced by the 
d'Alembertian, then $\beta=3/2$ is the optimal $\beta$ value for the corresponding 
``relativistic Vlasov-d'Alembert'' (rVdA) energy functional, and the
sharp value of the critical $\Lsp^{3/2}$ norm of $f$ can by computed.\footnote{Unpublished 
                        math. phys. seminar talk at the E. Schr\"odinger  
                        Inst., Vienna, Aug. 02, 2002. M.K. takes the opportunity to thank 
			N. Mauser for his kind invitation to present these results.}
        Curiously, the energy functional $\cE(f)$ of \rVPa\ is bounded below iff
the rVdA energy functional is, and by the same bound then, even though $\cE(f)$ could have been better 
behaved, a priori speaking, because in the rVdA energy functional $f$ and $\phi$ represent independent 
degrees of freedom.
        To exhibit these aspects very clearly, instead of elaborating on the approach of 
\cite{HadzicRein}, which is based on the Hardy--Littlewood--Sobolev inequality and standard 
interpolation arguments and which would also yield the sharp $\Lsp^{3/2}$ bound, 
we investigate the boundedness of $\cE(f)$ in terms of the topologically dual approach which 
can be applied almost verbatim also when the Poisson equation is replaced by the inhomogeneous 
wave equation.

	Thus we introduce the following functional of $f$ and $\phi$,
\beq
\wcE(f,\phi)
:=
\iint \Big(\sqrt{1+ |p|^2} +\phi (q) \Big) f(p,q) \dd{p}\,\dd{q}
+
\frac{1}{8\pi}\int \abs{\pdq\phi}^2 (q) \dd{q}
\,,
\label{EfuncFphi}
\eeq
irrespective of whether the functions $f$ and $\phi$ are related by the Poisson equation 
\refeq{rVPphiEQ} or not, and of whether $\phi$ satisfies \refeq{phiASqTOoo} or not.
	If $f$ and $\phi$ are related by $\phi = - |\id|^{-1}*\int f \dd{p}$,
as for solutions to \rVPa, then \refeq{EfuncF} and \refeq{EfuncFphi} can be converted into 
one another by using the Poisson equation \refeq{rVPphiEQ} and an integration by parts; in
those cases $\cE(f) = \min_{\psi\in \dot{\Hsp}^1_0}\wcE(f,\psi)$, and $\phi$ is the minimizer.
\begin{Prop}\label{propEboundOPT}
	Let 
	$\Omega = \{(f,\phi)\big| f\in\Psp_1\cap\Lsp^{3/2},\ \phi\in \dot{\Hsp}^1_0\}$.
	Then  
\beq
  \inf\left\{\wcE(f,\phi)\Big|(f,\phi)\in\Omega,\norm{f}_{3/2}\leq(3/8)(15/16)^{1/3}\right\} = 0.
\label{EboundFphi}
\eeq
	The $\Lsp^{3/2}$ bound on $f$ is sharp in the sense that, for any $\eps >0$, $C>0$, 
we can find $f\in \Psp_1\cap \Lsp^{3/2}$ with 
$\norm{f}_{3/2} = \frac{3}{8} \left(\frac{15}{16}\right)^{1/3}(1 + \eps)$,
and $\phi\in \dot{\Hsp}^1_0$ such that $\wcE \leq -C$.
	In particular, $\wcE$ is unbounded below if the $\Lsp^{3/2}$ bound on $f$ is
replaced by any $\Lsp^\beta$ bound for any $\beta <3/2$.
        Moreover, the infimum \refeq{EboundFphi} is not a minimum.
\end{Prop}
\newpage

\noindent
\textit{Proof of Proposition \ref{propEboundOPT}}.
	Using the estimates $|p| < \sqrt{1+|p|^2} \leq 1+|p|$, we have $\wcK < \wcE\leq 1+\wcK$, 
where
\beq
 \wcK(f,\phi)
  :=
    \frac{1}{8\pi} \int \abs{\pdq\phi}^2 (q) \dd{q}
   +
    \iint \Big( |p| +\phi (q) \Big) f(p,q) \dd{p}\,\dd{q}
\,,
\label{Kfunc}
\eeq
so that to prove the lower boundedness vs. unboundedness of $\wcE(f,\phi)$ as claimed,
it basically suffices to work with $\wcK(f,\phi)$. 
	Only for the precise value of the infimum do we need one extra estimate involving
$\sqrt{1+|p|^2}$.
	We also introduce the abbreviation $h(p,q) :=  |p| +\phi (q)$ 
for this auxiliary, ``ultra-relativistic'' single-particle Hamiltonian, and
we define $\cE_p^u(f):= \iint |p|f(p,q) \dd{p}\dd{q}$.

	To prove boundedness below of $\wcK(f,\phi)$ (hence, of $\wcE(f,\phi)$) on
the subset of $\Omega$ for which 
$\norm{f}_{3/2} \leq \frac{3}{8} \left(\frac{15}{16}\right)^{1/3}$, we begin by
noting that unboundedness below of $\wcK$ can only occur if $\phi(q) <0$ for some $q$; 
hence, we only need to show that the lesser functional $\underline{\wcK}\leq\wcK$ given by
\beq
 \underline{\wcK}(f,\phi)
   =
    \frac{1}{8\pi} \int \abs{\pdq\phi}^2 (q) \dd{q}
   -
    \iint h_-(p,q)f(p,q) \dd{p}\,\dd{q}
\label{KfuncAbound}
\eeq
is bounded below, where $h_- := -\min\{h,0\}\geq 0$ is the negative part of $h$.
	To this effect we now apply H\"older's inequality to $\int h_-f \dd{p}\dd{q}$, 
obtaining the estimate
\beq
 \underline{\wcK}(f,\phi)
  \geq
    \frac{1}{8\pi} \int \abs{\pdq\phi}^2 (q) \dd{q}
   - 
   \norm{h_-}_{\tau}\norm{f}_{\tau/(\tau-1)}
\,,
\label{KfuncBbound}
\eeq
with $\tau$ still to be determined.
	A simple integration with spherical coordinates in $p$ space gives
\beq
 \norm{h_-}_\tau^\tau
   =
 \frac{8\pi}{\prod_{k=1}^3(k+\tau)} {\norm{\phi_-}_{3+\tau}^{3+\tau}}
\label{hminusALPHAnorm}
\eeq
whenever $\norm{\phi_-}_{3+\tau}$ exists.
	Now the Sobolev embedding says\footnote{Note that we work with the
		homogeneous norm; the standard $\Hsp^1$ embedding of course gives
		${\Hsp}^1(\Rset^3) \to \Lsp^\alpha(\Rset^3)$ for all $\alpha\in [2,6]$.}
$\dot{\Hsp}^1_0(\Rset^3) \to \Lsp^\alpha(\Rset^3)$ iff $\alpha=6$.
        By \refeq{hminusALPHAnorm}, this means $\tau=3$, yielding $\tau/(\tau-1) = 3/2$ and
$\norm{h_-}_{3}^3 = ({\pi}/{15})  \norm{\phi_-}_6^6$.
	Using next the inclusion\footnote{Note that these fine details are necessary 
		only because we did not restrict $\phi$ by Poisson's equation \refeq{rVPphiEQ}
                with asymptotic condition \refeq{phiASqTOoo}; for if we had, then $\phi = -\phi_-$ 
                would ensue.}
$\supp\phi_-\subseteq\supp\phi$, we find the estimate
\beq
\norm{h_-}_{3}
\leq
\left(\frac{\pi}{15}\right)^{1/3}  \norm{\phi}_6^2
\,.
\label{hDREInormBOUND}
\eeq
	Hence, we set $\tau=3$ in \refeq{KfuncBbound} and use \refeq{hDREInormBOUND}, 
next recall the sharp Sobolev inequality for the embedding $\dot{\Hsp}^1_0 \to \Lsp^6$
(see \cite{Talenti, Lieb}), viz.
\beq
\norm{\pdq\phi}_2^2
-
3 \left({{\frac{\pi}{2}}}\right)^{4/3}\norm{\phi}_6^2
\geq
0
\,,
\label{SOBOLEVineq}
\eeq
where $3(\pi/2)^{4/3}$ is the largest possible coefficient for the $\norm{\phi}_6^2$ term, and
obtain
\beq
 \wcK(f,\phi)
  \geq
 \left(\frac{3}{8} \left(\frac{\pi}{16}\right)^{1/3} 
	-
 \left(\frac{\pi}{15}\right)^{1/3}\norm{f}_{3/2}\right)\norm{\phi}_6^2
\,.
\label{KfuncCbound}
\eeq
	Thus we have proved that
\beq
\inf\left\{\wcE(f,\phi)\Big|(f,\phi)\in\Omega,\norm{f}_{3/2}\leq(3/8)(15/16)^{1/3}\right\} \geq 0.
\eeq

	To see that $0$ is the best lower bound for $\wcE$ on $\Omega$ when 
$\norm{f}_{3/2} \leq \frac{3}{8} \left(\frac{15}{16}\right)^{1/3}$, 
we work with trial densities of the type
$f^\phi(p,q) := h_-^2(p,q)/\norm{h_-}_2^{2}$, with nonpositive $\phi\in\dot{\Hsp}^1_0$ 
still to be chosen.
	Clearly, $f^\phi\geq 0$ and $\norm{f^\phi}_{1}=1$, and furthermore$
\wcK(f^\phi,\phi) = \underline{\wcK}(f^\phi,\phi)$.
	We easily compute that
\beq
 \iint h(p,q) f^\phi(p,q) \dd{p}\dd{q}
  =
 -\frac{\norm{h_-}_3^3}{\norm{h_-}_2^2}
\label{TRIALhfA}
\eeq
and
$\norm{f^\phi}_{3/2}  = {\norm{h_-}_3^2}/{\norm{h_-}_2^2}$.
	With the help of \refeq{hminusALPHAnorm} and $\phi\leq 0$, we now find
$ \norm{h_-}_3^3  = ({\pi}/{15}) {\norm{\phi}_6^6}$,  and we readily calculate that
$ \norm{h_-}_2^2  = ({2\pi}/{15}) {\norm{\phi}_5^5}$,
so that the trial $\phi(q)$ will have to approach 0 sufficiently fast as $|q|\to\infty$
in order for $\norm{\phi}_5$ to exist; in particular, \refeq{phiASqTOoo} is fast enough.
	For such $f^\phi$, we therefore have
\beq
 \iint h(p,q) f^\phi(p,q) \dd{p}\dd{q}
  =
 - \frac{1}{2} \frac{\norm{\phi}_6^6}{\norm{\phi}_5^5}
\label{TRIALhfB}
\eeq
and
\beq
 \norm{f^\phi}_{3/2}
  =
 \frac{1}{2}\left(\frac{15}{\pi}\right)^{1/3}
 \frac{\norm{\phi}_6^4}{\norm{\phi}_5^5}
\,,
\label{TRIALfDREIHALBnormB}
\eeq
so that
\beq
\iint h(p,q) f^\phi(p,q) \dd{p}\dd{q}
=
- \left(\frac{\pi}{15}\right)^{1/3}\norm{f}_{3/2}\norm{\phi}_6^2.
\label{TRIALhfC}
\eeq
        By taking strong limits we now see that \refeq{TRIALhfC} is valid for
all $f\in\Psp_1\cap\Lsp^{3/2}$ and $\phi\in \Lsp^6$.
	In total we have $\wcK(f^\phi,\phi) = \cG(\phi)$,
with
\beq
\cG(\phi)
:= 	
{{\frac{1}{8\pi}}} \norm{\pdq\phi}_2^2
- \left({{\frac{\pi}{15}}}\right)^{1/3}\norm{f^\phi}_{3/2}\norm{\phi}_6^2
\,.
\label{Gfunc}
\eeq

	Now recall that up to translations, the scaling family of functions
\beq
\phi_\kappa(q)
=
- \frac{\kappa}{\sqrt{1 + \kappa^2 |q|^2}}
\label{eq:PlummerFCT}
\eeq
with $\kappa >0$ provides us with all the optimizers satisfying \refeq{phiASqTOoo} 
of the sharp Sobolev inequality \refeq{SOBOLEVineq} for the embedding $\dot{\Hsp}^1_0 \to \Lsp^6$,
i.e. ``$=$'' holds in \refeq{SOBOLEVineq} when $\phi=\phi_\kappa$; see \cite{Talenti, Lieb}.
        We set $\phi=\phi_\kappa$ in $f^\phi$; note that $f^{\phi_\kappa}\in\Psp_1\cap\Csp^\infty$,
so $f^{\phi_\kappa}\in \Psp_1\cap\Lsp^\alpha$ for all $\alpha$.
	An easy computation gives us
$\norm{\phi_\kappa}_6^4/\norm{\phi_\kappa}_5^5= 3\pi^{1/3}/4^{5/3}$, 
independent of $\kappa$.
	Hence, by \refeq{TRIALfDREIHALBnormB}, 
$\norm{f^{\phi_\kappa}}_{3/2} = \frac{3}{8}\left(\frac{15}{16}\right)^{1/3}$ independent of $\kappa$.
	Therefore $\cG\left(\phi_\kappa\right) =0$ 
for all $\kappa$, and since $\sup_{|p|\geq 0}\bigl( |p|\sqrt{1+ |p|^2} - |p|^2 \bigr)= 1/2$, 
we now conclude that for any small $\eps>0$ we have
\bea
 \wcE(f^{\phi_\kappa}, \phi_\kappa)
\!\!\!&=&\!\!\!
 \iint\! \left( \sqrt{1+ |p|^2} - |p| \right) f^{\phi_\kappa}(p,q) \dd{p}\,\dd{q}
\nonumber
\\
\!\!\!&\leq&\!\!\!
\frac{1}{2}\iint\!  |p|^{-1}f^{\phi_\kappa}(p,q) \dd{p}\,\dd{q}
\nonumber
\\
\!\!\!&=&\!\!\!
\frac{5}{4}
\frac{\norm{\phi_\kappa}_4^4}{\norm{\phi_\kappa}_5^5}
=
C_*\frac{1}{\kappa}
<
\eps
\eea
whenever $\kappa > C_*/\eps$, where $C_*$ is independent of $\kappa$.
	The ``$\inf$-part'' \refeq{EboundFphi} of our Proposition \ref{propEboundOPT} is proved.

        Incidentally, with a minor modification, the last line of reasoning also shows that 
``$\inf\neq\min$.''
        Thus, for any $(f,\phi)\in\Omega$ satisfying 
$\norm{f}_{3/2} \leq \frac{3}{8}\left(\frac{15}{16}\right)^{1/3}$, we have
\bea
 \wcE(f, \phi)
\!\!\!&=&\!\!\!
 \iint\! \left( \sqrt{1+ |p|^2} - |p| \right) f(p,q) \dd{p}\,\dd{q}
+
 \wcK(f,\phi)
\nonumber
\\
\!\!\!&\geq&\!\!\!
 \iint\! \left( \sqrt{1+ |p|^2} - |p| \right) f(p,q) \dd{p}\,\dd{q} \ > 0
\,.
\label{infNOTmin}
\eea

	Next, to prove that the $\Lsp^{3/2}$ bound on $f$ is sharp, we now show that for any given 
$\eps >0$ and $C>0$, we can find $f\in \Psp_1\cap \Lsp^{3/2}$ and $\phi\in \dot{\Hsp}^1_0$ 
such that $\norm{f}_{3/2}= \frac{3}{8}\left(\frac{15}{16}\right)^{1/3}(1 + \eps)$ 
and $\wcE \leq -C$.
	As stated at the beginning of the proof of our proposition, it suffices to
prove unboundedness below for $\wcK$ along these lines.
	For this purpose we continue to work with the optimizers of Sobolev's inequality 
\refeq{SOBOLEVineq}, $\phi_\kappa$, but now invoke the double scaling family of trial densities 
$f= f^{\phi_1}_{\kappa,\lambda}$, $\kappa>0,\lambda>0$, defined by
\beq
f^{\phi_1}_{\kappa,\lambda}(p,q):= \kappa^3\lambda^3 f^{\phi_1}(\lambda{p},\kappa{q})
\,.
\label{doubleSCALINGf}
\eeq
        Note that $f^{\phi_1}_{\kappa,\lambda}\in\Psp_1\cap\Csp^\infty$ for all 
$\kappa,\lambda$, so $f^{\phi_1}_{\kappa,\lambda}\in  \Psp_1\cap\Lsp^\alpha$ for all $\alpha$. 
        Note furthermore by an obvious re-scaling of the integration variable that the
pertinent density in $q$ space is independent of $\lambda$, i.e. 
$\int f^{\phi_1}_{\kappa,\lambda}(p,q)\dd{p} = \int f^{\phi_1}_{\kappa,1}(p,q)\dd{p}$ for all 
$\lambda$; moreover, it equals $\int f^{\phi_\kappa}(p,q)\dd{p}=:\rho^{\phi_\kappa}(q)$, for
$f^{\phi_1}_{\kappa,1/\kappa} = f^{\phi_\kappa}$.
         The density $\rho^{\phi_\kappa}(q)=\kappa^3\rho^{\phi_1}(\kappa q)$, where $\rho^{\phi_1}$
is given by 
\beq
 \rho^{\phi_1}(q)
=
\frac{3}{4\pi} (-\phi_1)^5(q)
\label{RHOkappa}
\,.
\eeq
        Thus, not only are $f^{\phi_\kappa}(p,q)$ and $\phi_\kappa(q)$ related by Poisson's equation 
\refeq{rVPphiEQ} with \refeq{phiASqTOoo}, but so are $f^{\phi_1}_{\kappa,\lambda}(p,q)$ and 
$\phi_\kappa(q)$ for all $\lambda$.
        Next, a simple computation yields
\beq
\big\|{f^{\phi_1}_{\kappa,\lambda}}\big\|_{3/2}
=
\kappa\lambda \norm{f^{\phi_1}}_{3/2}
\,,
\label{fkappalambdaDREIHALBEnorm}
\eeq
with $\norm{f^{\phi_1}}_{3/2} = \frac{3}{8} \left(\frac{15}{16}\right)^{1/3}$, so setting 
$\big\|f^{\phi_1}_{\kappa,\lambda}\big\|_{3/2} = \frac{3}{8} \left(\frac{15}{16}\right)^{1/3}(1 + \eps)$
in \refeq{fkappalambdaDREIHALBEnorm} defines a branch of a hyperbola $\kappa\lambda = 1+\eps$ 
in the first quadrant of Cartesian $(\kappa,\lambda)$-parameter space.
         Along this hyperbola branch, $f^{\phi_1}_{\kappa,\lambda}$ is a probability density with 
$\Lsp^{3/2}$ norm equal to $(1+\eps)C_{3/2}$, i.e. bigger than the acclaimed critical $C_{3/2}$.
         Next, a straightforward computation shows that along this hyperbola branch,
\beq
\wcK\bigl(f^{\phi_1}_{\kappa,(1+\eps)/\kappa},\phi_\kappa\bigr)
=
- \frac{3\pi}{32}\frac{\eps}{1+\eps}\kappa < 0
\eeq
for all $\kappa>0$. 
	Hence, for any $\eps >0$ and $C >0$ there is a 
unique $\kappa(\eps,C) = ({32}/{3\pi})(1+1/\eps)C > 0$,   
such that $\wcK\bigl(f^{\phi_1}_{\kappa,(1+\eps)/\kappa},\phi_\kappa\bigr)< - C$
whenever $\kappa > \kappa(\eps,C)$.
	Therefore, $\wcK\bigl(f^{\phi_1}_{\kappa,(1+\eps)/\kappa},\phi_\kappa\bigr)$
is unbounded below, and this implies unboundedness below of $\wcK(f,\phi)$ over the
set $\{(f,\phi)\in\Omega,\, \norm{f}_{3/2} \leq C\}$ when $C> \frac{3}{8}\left(\frac{15}{16}\right)^{1/3}$;
hence, the same holds for $\wcE(f,\phi)\ [\leq 1 +\wcK(f,\phi)]$.

        Finally we prove that $\wcK(f,\phi)$ is unbounded below on the domain
$\{\phi\in\dot{\Hsp}^1_0, f\in \Psp_1\cap \Lsp^\beta, \norm{f}_\beta \leq C_\beta\}$ for 
any $\beta \in (1,3/2)$ and any $C_\beta>0$. 
        (Notice that for $\beta=1$, which only allows $\norm{f}_1=1$, we already proved unboundedness 
at the beginning of section 3.)
        We ignore \refeq{rVPphiEQ}, as allowed by the hypotheses of Proposition \ref{propEboundOPT}.
        Thus, since for $\beta<3/2$ the interval $(2,3/\beta)$ is not empty,
we pick any $\vartheta\in (2,3/\beta)$.
	We now choose a family of trial densities of the type
$\hat{f}^\phi(p,q): = h_-^\vartheta(p,q)/\norm{h_-}_\vartheta^{\vartheta}$, 
with some nonpositive $\phi\in\dot{\Hsp}^1_0$ satisfying \refeq{phiASqTOoo}; $\phi$
will be specified  further below.
	Also for this trial family of $f$s we have
$\wcK(\hat{f}^\phi,\phi) = \underline{\wcK}(\hat{f}^\phi,\phi)$.
	Moreover, using again \refeq{hminusALPHAnorm}, we find
\beq
 \big\|\hat{f}^\phi\big\|_\beta
  =
 a(\vartheta,\beta)
 \frac{\norm{\phi}_{\vartheta\beta+3}^{\vartheta+3/\beta}}{\norm{\phi}_{\vartheta+3}^{\vartheta+3}}
\,,
\label{TRIALfBETAnorm}
\eeq
and
\beq
\wcK(\hat{f}^\phi,\phi) =
\frac{1}{8\pi}\norm{\pdq\phi}_2^2
 -
 b(\vartheta)
 \frac{\norm{\phi}_{\vartheta+4}^{\vartheta+4}}{\norm{\phi}_{\vartheta+3}^{\vartheta+3}}
\,,
\eeq
where $a$ and $b$ are some numerical constants dependent on the displayed arguments.
	Now notice that for the stipulated range of $\beta$ and $\vartheta$ values, 
we have $5<\vartheta+3 < 6$ and $5< \vartheta\beta +3 < 6$, but $\vartheta +4 > 6$. 
	Since for $\phi\in\dot{\Hsp}^1_0$ satisfying \refeq{phiASqTOoo} we actually 
have $\phi \in \Lsp^\alpha(\Rset^3)$ for all $\alpha\in (3,6]$, it follows right away
that $\big\|\hat{f}^\phi\big\|_{\beta}<\infty$. 
	But among these $\phi$ functions there are infinitely many for which 
$\norm{\phi}_{\vartheta+4}^{\vartheta+4}=\infty$.  
	Let $\phi_*$ be such a $\phi$. 
	We then can make
$\big\|\hat{f}^{\phi_*}\big\|_{\beta}\leq C_\beta$ for any small $C_\beta$
by multiplying $\phi_*$ by some small constant $c$ if necessary (note that $c\phi_*$
no longer satisfies \refeq{phiASqTOoo}, though), while  
$\norm{\phi_*}_{\vartheta+4}^{\vartheta+4}=\infty$ implies
$\wcK(\hat{f}^{\phi_*},\phi_*) = - \infty$.
	Therefore, we obtain lower unboundedness of $\wcE(f,\phi)$ 
on the domain
$\{\phi\in\dot{\Hsp}^1_0, f\in \Psp_1\cap \Lsp^\beta, \norm{f}_\beta \leq C_\beta\}$
for any $1<\beta<3/2$ and any $C_\beta$.
	So $\Lsp^{3/2}$ in \refeq{EboundFphi} cannot be replaced by any 
$\Lsp^\beta$ with $\beta< 3/2$.
\QED
\begin{Rema}
	Our proof, a variant of which was  announced a while ago (see footnote 4),
was inspired by a related proof in \cite{Aly}  (which in turn was inspired by one in \cite{WieZieSch})
for the energy functional of the nonrelativistic VP system with $\sigma=-1$.
	For VP$^-$ the critical $\beta ={9/7}$, and the critical $\Lsp^{9/7}$
bound on $f$ simply reads $\norm{f}_{9/7}<\infty$.
\end{Rema}
        We are now ready to address the boundedness of $\cE(f)$.
\begin{Prop}\label{propEofFbound}
	We have
\beq
\inf\left\{\cE(f)\Big|f\in \Psp_1\cap \Lsp^{3/2},\
	   \norm{f}_{3/2}\leq (3/8)(15/16)^{1/3}
   \right\} = 0
\,,
\label{Ebound}
\eeq
and the infimum \refeq{Ebound} is not a minimum.
        Moreover, we have 
\beq
\inf\left\{\cE(f)\Big|f\in \Psp_1\cap \Lsp^\beta,\
	   \norm{f}_{\beta}\leq C\right\} = - \infty
\label{noEbound}
\eeq
whenever $C> C_\beta$, with 
\beq
    C_\beta : = 
\inf_{\Psp_1\cap\Lsp^\beta} \left(\frac{\cE_p^u(f)}{-\cE_q(f)}\right)^{3(1-1/\beta)}
\norm{f}_{\beta}\, .
\label{CbetaDEF}
\eeq 
        In particular, $C_\beta = 0$ for all $\beta\in (1,3/2)$.
        Furthermore, for all $\beta\geq 3/2$ we have 
\beq
\left[\left(\frac{3}{8}\right)^3\frac{15}{16}\right]^{1-1/\beta} \leq  C_\beta \leq 
    \frac{45}{8\pi^2}
\left( \frac{8\pi^{5/2}}{\prod_{k=1}^3(k+2\beta)}\frac{\Gamma(\beta)}{\Gamma(\beta+\frac{3}{2})}
\right)^{1/\beta}.
\label{CbetaBOUND}
\eeq 
        At $\beta=3/2$ the upper and lower bounds in \refeq{CbetaBOUND} coincide; 
$C_{3/2}\! =\! (3/8)(15/16)^{1/3}$ is the sharp value, i.e. the bound on the $\Lsp^{3/2}$ 
norm of $f$ in \refeq{Ebound} is optimal.
\end{Prop}

\epsfxsize=7.5cm
\centerline{\epsffile{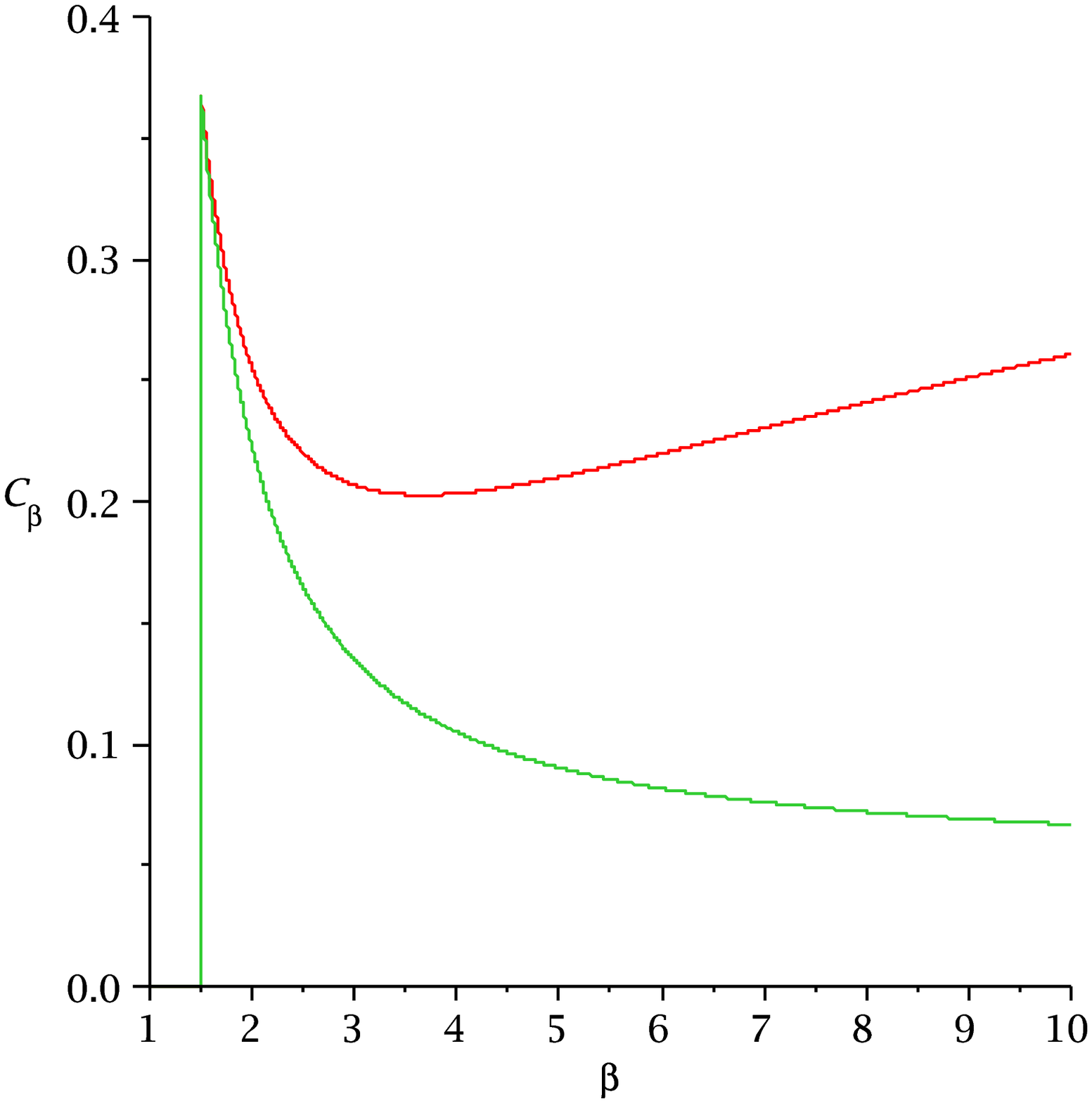}}
\begin{center}
{\scriptsize{
Upper and lower bounds on $C_\beta$ together with the vertical line segment between $0$ and $C_{3/2}$.
\\
Replacing the co-ordinate label $C_\beta$ by $\|f\|_\beta$, the diagram acquires the following meaning:\\
On and to the right of the vertical line segment, yet on and below the lower curve, we have\ \\
$\inf\cE(f)\geq 0$, while $\inf \cE(f)\!=\! -\infty$ to the left of the vertical line and above 
the upper curve.\qquad \\ }}
\end{center}
\newpage

\noindent
\textit{Proof of Proposition \ref{propEofFbound}}.
        As to \refeq{Ebound}, since the Poisson equation \refeq{rVPphiEQ} together with \refeq{phiASqTOoo} 
selects a subset of $\Omega$, it follows that $\cE(f)$ is bounded below whenever $\wcE(f,\phi)$ is.
	Furthermore, since we proved the infimum versus unboundedness of $\wcE(f,\phi)$
when $\phi\in\dot{\Hsp}^1_0$ and $f\in \Psp_1\cap \Lsp^{3/2}$ with pairs of trial functions 
$(f^{\phi_\kappa},\phi_\kappa)$ respectively $(f^{\phi_1}_{\kappa,\lambda},\phi_\kappa)$
(with $\phi_\kappa$ given by \refeq{eq:PlummerFCT}),
either pair of which solves Poisson's equation \refeq{rVPphiEQ} together with \refeq{phiASqTOoo},
the same infimum versus unboundedness features hold for $\cE(f)$ when
$f\in \Psp_1\cap \Lsp^{3/2}$; viz. \refeq{Ebound} is optimal, indeed, and ``$\inf\neq\min$.''

	As to \refeq{noEbound} and \refeq{CbetaDEF}, we set $\cK(f): = \cE_p^u(f) + \cE_q(f)$. 
	As with $\wcK(f,\phi)$ and $\wcE(f,\phi)$, unboundedness below of $\cK(f)$ implies unboundedness 
below of $\cE(f)$.
	In this vein, by double scaling we find $\cK(f_{\kappa,\lambda})=\lambda^{-1}\cE^u_p(f)+\kappa\cE_q(f)$,
so for any $f$ we can give $\cK(f_{\kappa,\lambda})$ any value we like by choosing $\kappa$ and $\lambda$ 
appropriately, at the expense of changing 
$\norm{f}_\beta$ to $\norm{f_{\kappa,\lambda}}_\beta = (\kappa\lambda)^{3(1-1/\beta)}\norm{f}_\beta$.
	In particular, once we found a special $(\kappa_0,\lambda_0)$ pair for which 
$\cK(f_{\kappa_0,\lambda_0})<0$, 
we can then let $\cK(f_{\kappa,\lambda})\downarrow -\infty$ by scaling along the branch of a hyperbola 
$\kappa\lambda = \kappa_0\lambda_0$ in the first quadrant of Cartesian $(\kappa,\lambda)$-parameter space, 
keeping $\norm{f_{\kappa,\lambda}}_\beta$ fixed along this scaling sequence. 
	For each $f$, the borderline case $\cK(f_{\kappa,\lambda}) =0$ is obtained by choosing 
$\kappa\lambda = -\cE^u_p(f)/\cE_q(f)$. 
	Hence, \refeq{noEbound} holds whenever $C>C_\beta$, with $C_\beta$ the infimum over 
$f\in\Psp_1\cap\Lsp^\beta$ of 
$\norm{f_{\kappa,\lambda}}_\beta = (-\cE^u_p(f)/\cE_q(f))^{3(1-1/\beta)}\norm{f}_\beta$, 
i.e. \refeq{CbetaDEF}.

	As for \refeq{CbetaBOUND}, since $(f^{\phi_1}_{\kappa,\lambda},\phi_\kappa)$ jointly solves Poisson's
equation \refeq{rVPphiEQ} with \refeq{phiASqTOoo}, our proof of unboundedness of 
$\wcE(f,\phi)$ when $\big\|f^{\phi_1}_{\kappa,\lambda}\big\|_{3/2} 
	= \frac{3}{8} \left(\frac{15}{16}\right)^{1/3}(1 + \eps)$
applies nearly verbatim, with $\big\|f^{\phi_1}_{\kappa,\lambda}\big\|_{3/2}$ 
now replaced by $\big\|f^{\phi_1}_{\kappa,\lambda}\big\|_{\beta}$
and $\frac{3}{8} \left(\frac{15}{16}\right)^{1/3} (=\norm{f^{\phi_1}}_{3/2}=C_{3/2})$ replaced by
$\norm{f^{\phi_1}}_{\beta} =$ r.h.s.\refeq{CbetaBOUND}, giving \refeq{noEbound} whenever 
$C>$ r.h.s.\refeq{CbetaBOUND}, implying the second inequality in \refeq{CbetaBOUND}.
       The first inequality in \refeq{CbetaBOUND} follows from the interpolation inequality and \refeq{Ebound}, 
cf. our Remark 1.4.

	Lastly, to prove $\inf\cE(f)=-\infty$ when $\beta\in (1,3/2)$, we use
$\cE(f)= \min_{\phi\in \dot{\Hsp}^1_0}\wcE(f,\phi)$ and the fact that 
$\inf_\Omega\wcE(f,\phi)=-\infty$ for this $\beta$ interval (for which proof we conveniently ignored 
\refeq{rVPphiEQ} and \refeq{phiASqTOoo}). 
	Yet, it is desirable to also have a direct proof for $\cE(f)$.

	We need to distinguish between $\beta\in (1,6/5)$ and $\beta\in [6/5,3/2)$.
	First, when $\beta< 6/5$, then by taking product densities $f(p,q)=g(p)\rho(q)$ we have
$\rho\in\Lsp^\beta$ for $\beta<6/5$, and then it is well-known that we can find $f$ with 
$\cE_p^u(f)<\infty$ and $\norm{f}_\beta<\infty$ but with $\cE_q(f)=-\infty$. 
        Hence $C_\beta = 0$ when $\beta\in (1,6/5)$.

	Next, when $\beta\in [6/5,3/2)$, we can recycle the trial densities of the type 
$\hat{f}^\phi(p,q) = h_-^\vartheta(p,q)/\norm{h_-}_\vartheta^{\vartheta}$, 
introduced in the proof of Proposition \ref{propEboundOPT}, but now with $\phi(q)$ replaced by 
$\psi_\delta(q) = - e^{-|q|}/|q|^\delta$, where $\delta>0$ will be specified below; also
$\vartheta >0$ will need to be re-specified as well.
	In particular, $\psi_\delta$ here is \emph{not} the self-consistent potential $\phi$
of \rVPa, which $\phi$ is related to $\hat{f}$ by Poisson's equation.
	Rather, $\psi_\delta$ means a convenient way of generating a suitable family of densities 
$\rho=\hat{\rho}^{\psi_\delta}  =
{(-\psi_\delta)^{\vartheta+3}}/{\norm{\psi_\delta}_{\vartheta+3}^{\vartheta+3}}$
which exhibit a local power law singularity; the exponential factor serves to avoid integrability
problems at spatial infinity that could occur with these power laws.
        We will show that for $\beta\in [6/5,3/2)$ we can find $\delta>0$ and $\vartheta>0$
such that $\hat{f}^{\psi_\delta}\in \Lsp^{\beta}$ but $\hat{f}^{\psi_\delta}\not\in \Lsp^{3/2}$, 
and such that $\cE_p^u(\hat{f}^{\psi_\delta}) <\infty$ while $\cE_q(f) =-\infty$.

        Indeed, since $\big\|\hat{f}^{\psi_\delta}\big\|_\beta$ is given by \refeq{TRIALfBETAnorm} 
with $\phi$ replaced by $\psi_\delta$, we have $\hat{f}^{\psi_\delta}\not\in \Lsp^{3/2}$ yet 
$\hat{f}^{\psi_\delta}\in \Lsp^\beta$ for any given $\beta\in[6/5,3/2)$ 
whenever $\delta \in [2/(2+\vartheta), 3/(3 +\vartheta\beta))$, which
interval is not empty when $\vartheta >0$.
       For the just established range of $\delta$ values, given any $\beta\in [6/5,3/2)$ 
and for any $\vartheta >0$, we also have $\hat{\rho}^{\psi_\delta}\in \Psp_1\cap\Lsp^1$. 
       (All of the above holds also with $\beta\in(1,3/2)$.)
       Next, since $\rho\in\Lsp^{6/5}$ implies $\cE_q(f)>-\infty$, we certainly want 
$\hat{\rho}^{\psi_\delta}\not\in\Lsp^{6/5}$, which means we want
$
\norm{\hat{\rho}^{\psi_\delta}}_{6/5}
  =
{\norm{\psi_\delta}^{(\vartheta+3)}_{6(\vartheta+3)/5}}/{\norm{\psi_\delta}_{\vartheta+3}^{\vartheta+3}}
$
to be $\infty$.
        This gives us the condition $\delta \in[ 5/(6+2\vartheta), 3/(3+\vartheta))$, which interval
is not empty for any $\vartheta>0$.
        Taking the intersection of our two $\delta$ intervals we find the common condition
$\delta \in [\max\{2/(2+\vartheta), 5/(6+2\vartheta)\}, 3/(3 +\vartheta\beta))$, which 
interval is not empty for $\beta\in [6/5,3/2)$ when $\vartheta \in (0,3/(5\beta-6))$; note that
$5/(6+2\vartheta) <2/(2 +\vartheta)$ for $\vartheta\in(0,2)$ and $5/(6+2\vartheta) > 2/(2 +\vartheta)$  
for $\vartheta >2$.
	Lastly, we want that $\cE_p^u(\hat{f}^{\psi_\delta}) <\infty$.
        Since 
$\cE_p^u(\hat{f}^{\psi_\delta}) 
=
b(\vartheta) {\norm{\psi_\delta}_{\vartheta+4}^{\vartheta+4}}/{\norm{\psi_\delta}_{\vartheta+3}^{\vartheta+3}}$,
the necessary and sufficient condition for $\cE_p^u(\hat{f}^{\psi_\delta}) <\infty$ is
$\delta < 3/(4+\vartheta)$, which may imply, or be implied by, $\delta < 3/(3+\vartheta\beta)$, 
depending on $\beta$; moreover, $3/(4+\vartheta)> \max\{2/(2+\vartheta), 5/(6+2\vartheta)\}$ iff 
$\vartheta > 2$, but in that case $\max\{2/(2+\vartheta), 5/(6+2\vartheta)\} =5/(6+2\vartheta)$.
        Hence, the resulting common intersection of all our $\delta$ intervals is
the interval $[5/(6 +2\vartheta), \min\{3/(3+\vartheta \beta),3/(4+\vartheta)\})$, which 
is not empty for $\beta \in [6/5,3/2)$ when $\vartheta\in (2, 3/(5\beta-6))$.
        In summary, for $\beta \in [6/5,3/2)$, when $\vartheta \in (2, 3/(5 \beta -6))$
and $\delta \in [5/(6 +2\vartheta), \min\{3/(3+\vartheta \beta),3/(4+\vartheta)\})$, then we have that
$\hat{f}^{\psi_\delta} \in \Lsp^\beta$ with $\beta \in [6/5, 3/2)$ but 
$\hat{f}^{\psi_\delta} \not\in \Lsp^{3/2}$; we also have $\cE^u_p(\hat{f}^{\psi_\delta}) <\infty$;
furthermore, the pertinent $\hat\rho^{\psi_\delta} \not\in \Lsp^{6/5}$.
	A brief calculation shows for $\hat\phi(|q|)= - |\id|^{-1}*\int\! \hat{f}^{\psi_\delta}(p,q)\dd{p}$
that $\hat\phi^\prime(|q|) \asymp |q|^{1-\delta(3+\vartheta)}$ near the singularity, so 
$\hat\phi\not\in \dot{H}^1$ when $\delta \geq 3/(6+2\vartheta)$. 
	Thus, $\delta \geq 5/(6+2\vartheta)$ implies $\cE_q(\hat{f}^{\psi_\delta})=-\infty$, and so
\beq
	C_\beta =
\biggl(\frac{\cE_p^u(\hat{f}^{\psi_\delta})}{-\cE_q(\hat{f}^{\psi_\delta})}\biggr)^{3(1-1/\beta)}
\big\|\hat{f}^{\psi_\delta}\big\|_{\beta} 
	= 0 \quad \mathrm{when}\quad \beta\in [6/5,3/2).
\eeq
	Indeed, $\cK(\hat{f}^{\psi_\delta}) =-\infty$, which implies $\cE(\hat{f}^{\psi_\delta}) = -\infty$.\QED
\newpage
\begin{Defi}
	In the following, we will call initial data $f_0\in \Psp_1\cap \Lsp^{3/2}$
{\textbf{subcritical}} if $\norm{f_0}_{3/2} < C_{3/2}$, 
{\textbf{critical}} if $\norm{f_0}_{3/2}= C_{3/2}$, 
and 
{\textbf{supercritical}} if $\norm{f_0}_{3/2} > C_{3/2}$. 
        We use the analogous terminology for the solutions launched by such data.
\end{Defi}
\section{A-priori bounds for subcritical $f$}
        By \refeq{infNOTmin}, for all critical and subcritical $f$ we have that
$\cE(f)=\wcE(f,\phi[f])>0$, where
\beq
 \phi[f]= - |\id|^{-1}*\int f \dd{p}
\label{phiOFf}
\,.
\eeq
        Unfortunately, this does not seem to lend itself to further estimates on the individual
energy contributions $\cE_p(f)$ and $\cE_q(f)$.
        Yet, for subcritical $f$ we actually have a better result than \refeq{infNOTmin}, and this 
does lead to a-priori bounds on  $\cE_p(f)$ and $\cE_q(f)$.
\begin{Prop}\label{propEboundSUBcrit}
	For any $f\in \Psp_1\cap \Lsp^{3/2}$ satisfying $\norm{f}_{3/2} < C_{3/2}$, we have 
\beq
 \cE(f)
  \geq
 \frac{1}{8\pi \varkappa(f)} \int \abs{\pdq\phi[f]}^2(q) \dd{q}
\label{Edominance}
\eeq
with
\beq
\varkappa(f) = \frac{C_{3/2}}{C_{3/2}-\norm{f}_{3/2}}
\,.
\eeq
\end{Prop}
\noindent
\textit{Proof of Proposition \ref{propEboundSUBcrit}.}
	Since by hypothesis $\norm{f}_{3/2}$ is strictly less than the
sharp critical value, we can retain a little bit from the Dirichlet integral and see, 
by inspecting the steps of the proof of the lower boundedness of $\wcE$, hence of $\cE$, 
that now we get the estimate \refeq{Edominance}.
\QED

	Proposition \ref{propEboundSUBcrit} has the following important spin-off.
\begin{Coro}\label{coroBOUNDSf}
	Let $f\in\Psp_1\cap\Lsp^{3/2}$ satisfy 
$\norm{f}_{3/2}<\frac{3}{8} \left(\frac{15}{16}\right)^{1/3}$.
	Then, 
\bea
 \norm{\pdq\phi[f]}_2^2 
&\leq &
8\pi\varkappa(f) \cE(f)
\,,
\label{coroBOUNDSfEpot}
\\
\iint \sqrt{1+|p|^2} f(p,q)\dd{p}\dd{q} 
&\leq&
 (1+\varkappa(f))\cE(f)
\,.
\label{coroBOUNDSfEkin}
\eea
\end{Coro}

\noindent\textit{Proof of Corollary \ref{coroBOUNDSf}.}
	The bound \refeq{coroBOUNDSfEpot} on $\norm{\pdq\phi}_2$ is just a restatement of
\refeq{Edominance} in Proposition \ref{propEboundSUBcrit}.

	Noting that for $\phi$ given by \refeq{phiOFf} the energy functional \refeq{EfuncF} 
can be rewritten as
\beq
 \cE(f) 
  =
   \iint \sqrt{1+ |p|^2} f(p,q) \dd{p}\,\dd{q}
  -
  \frac{1}{8\pi}\norm{\pdq\phi[f]}_2^2
\,,
\label{EfuncFandPHI}
\eeq
we see right away that \refeq{coroBOUNDSfEpot} now implies \refeq{coroBOUNDSfEkin}.
\QED

	To state our next corollary we need the following lemma, which does not 
explicitly require $f$ to be subcritical.
\begin{Lemm}\label{lemmRHObound}
	Assume that $f\in\Psp_1\cap\Lsp^\alpha$ for some $\alpha\geq 1$ 
(with $\alpha=\infty$ allowed).
	Then there exists a $C(\alpha)$ that depends only on $\alpha$, such that the 
relative density of particles (given in \refeq{rhoASfINT}) satisfies the bound
\beq
\norm{\rho}_{\gamma} 
\leq
C(\alpha) \|f\|_\alpha^{\eta}\cE_{p}(f)^{1-\eta},
\label{rhoLbetaBOUND}
\eeq
where (we recall that) $\cE_p(f)$ is the kinetic plus rest energy of $f$ 
(i.e., l.h.s.\refeq{coroBOUNDSfEkin}), and 
\beq
\gamma := \frac{4\alpha-3}{3\alpha-2}, \qquad \eta := \frac{\alpha}{4\alpha-3}
\label{betaANDeta}
\eeq
\end{Lemm}
\begin{Rema}
	Note that $\eta$ in \refeq{betaANDeta} is a decreasing function of $\alpha$,
taking values in $[1/4,1]$, while $\gamma$ in \refeq{betaANDeta} is an increasing 
function of $\alpha$, taking values in $[1,4/3]$. 
	So the optimal possible control of $\rho$ is with exponent $\gamma= 4/3$, obtained
when $\alpha = \infty$, while any $\alpha <\infty$ necessarily entails a weaker control on $\rho$.
	In particular, when $\alpha=1$ nothing new is learned beyond what the definition of 
$\rho$ says already.
\end{Rema}
\noindent
\textit{Proof of Lemma \ref{lemmRHObound}.}
        Inspection of the proof, in \cite{HorstA} and \cite{GlasseySchaefferA},
of the corresponding $\norm{\rho}_{4/3}$ bound for the relativistic Vlasov--Maxwell equations 
when $f\in\Psp_1\cap\Lsp^\infty$ is assumed shows that their proof generalizes to
$f\in\Psp_1\cap\Lsp^\alpha$, all $\alpha$.
        Thus,
\begin{eqnarray*}
\rho(q)\!\! & =&\!\! \int_{|p|<P} f \dd{p} + \int_{|p|\geq P} f \dd{p} \\
& \leq & \Big(\int |f|^\alpha \dd{p}\Big)^{1/\alpha} 
({\textstyle{\frac{4\pi}{3}}}P^3)^{1/\alpha'} 
+ \frac{1}{P} \int \sqrt{1+|p|^2} f \dd{p} \\
& := & ({\textstyle{\frac{4\pi}{3}}})^{1/\alpha'} F(f) P^{3/\alpha'} + G(f) P^{-1}
\end{eqnarray*}
which upon optimizing in $P$ yields
\beq
 \rho(q) \leq C(\alpha) (F(f))^\eta (G(f))^{1-\eta},\qquad \eta:= {\alpha}/{(4\alpha-3)}.
\eeq
	Raising both sides to power $\gamma$, integrating in $q$ and applying H\"older's inequality 
with exponent $\delta :=  3\alpha-2 = \alpha/(\gamma\eta)$, noting that $\gamma(1-\eta)\delta' = 1$, 
yields
\begin{eqnarray*} 
\|\rho\|_\gamma^\gamma 
& \leq & 
C(\alpha)^\gamma \int (F(f))^{\eta\gamma}(G(f))^{(1-\eta)\gamma} \ \dd{q}
\\
& \leq & 
C(\alpha)^\gamma \Big(\int (F(f))^{\eta\gamma\delta}\dd{q}\Big)^{1/\delta} 
\Big( \int (G(f))^{(1-\eta)\gamma\delta'}\dd{q} \Big)^{1/\delta'} 
\\
& = & 
C(\alpha)^\gamma
\|f\|_\alpha^{\alpha/\delta} \Big(\iint \sqrt{1+|p|^2}f(p,q)\dd{p}\,\dd{q}\Big)^{\gamma(1-\eta)}
\end{eqnarray*}
which gives the desired bound by virtue of the hypotheses of Lemma \ref{lemmRHObound}.\QED

	As a spin-off of Corollary \ref{coroBOUNDSf} and Lemma \ref{lemmRHObound} we now have
\begin{Coro}\label{coroRHOsubCRITbound}
	Let $f\in\Psp_1\cap\Lsp^{\alpha}$ for some $\alpha\geq 3/2$.
	Assume furthermore that $\norm{f}_{3/2}<\frac{3}{8} \left(\frac{15}{16}\right)^{1/3}$. 
	Then there exists a $C(\alpha)$ that depends only on $\alpha$,
such that the relative density of particles satisfies the bound
\beq
\norm{\rho}_{\gamma} 
\leq 
C(\alpha) \|f\|_\alpha^{\eta} (1+\varkappa(f))^{1-\eta}\cE(f)^{1-\eta},
\label{rhoSUBcritLbetaBOUND}
\eeq
with $\gamma$ and $\eta$ related to $\alpha$ by \refeq{betaANDeta}.
\end{Coro}

\noindent
\textit{Proof of Corollary \ref{coroRHOsubCRITbound}.}
	Under the hypotheses of Corollary \ref{coroRHOsubCRITbound} we can apply Corollary 
\ref{coroBOUNDSf} which asserts in \refeq{coroBOUNDSfEkin} that $\cE_{p}(f)\leq(1+\varkappa(f))\cE(f)$, 
which bound gives the desired bound on $\norm{\rho}_\gamma$ by virtue of \refeq{rhoLbetaBOUND}.\QED
\begin{Rema}
	Note that in \refeq{rhoSUBcritLbetaBOUND} $\gamma$ now takes a value in $[6/5,4/3]$, 
while $\eta$ takes a value in $[1/4,1/2]$.
       For the smallest possible $\alpha= 3/2$, \refeq{rhoSUBcritLbetaBOUND}
yields only an $\Lsp^{6/5}$ estimate of $\rho$, viz.\footnote{Incidentally, for $\vartheta\in(0,2)$ 
                 and $\delta\in [5/(6+2\vartheta), 2/(2+\vartheta))$, our
                 $\hat{f}^{\psi_\delta}\propto (|p|-e^{-|q|}/|q|^{\delta})_-^\vartheta \in\Lsp^{3/2}$ but
                 $\hat{\rho}^{\psi_\delta}\not\in\Lsp^{6/5}$.
                 Yet we also have $\cE_p^u(\hat{f}^{\psi_\delta}) = \infty$
                 for all $\delta\in [5/(6+2\vartheta), 2/(2+\vartheta)]$ when $\vartheta\in(0,2]$.}
\beq
\hskip 3truecm
\norm{\rho}_{6/5}
\leq
C \|f\|_{3/2}^{1/2} (1+\varkappa(f))^{1/2}\cE(f)^{1/2},
\label{rhoSUBcritL1pt5BOUND}
\eeq
where $C$ is some numerical constant independent of $f$.
      Of course, even weaker $\Lsp^{\gamma^\pr}$ estimates of $\rho$ hold, since under the hypotheses of 
Corollary \ref{coroRHOsubCRITbound}, \refeq{rhoSUBcritLbetaBOUND} remains true if in 
\refeq{rhoSUBcritLbetaBOUND} $\alpha$ is replaced by any $\alpha^\pr\in[1,\alpha)$, 
with $\gamma\mapsto\gamma^\pr$ and $\eta\mapsto\eta^\pr$ correspondingly.
\end{Rema}
\section{Subcritical solutions}
	We note that the a-priori bounds on the kinetic and potential energy functionals
and on the $\Lsp^\gamma$ norm of $\rho$ for subcritical $f$ depend on $f$ only through
functionals which are conserved by sufficiently integrable classical solutions of the Vlasov evolution.
	This implies at once uniform bounds w.r.t. time on the corresponding quantities for
sufficiently integrable classical solutions.
\begin{Coro}\label{coroBOUNDSfOFt}
	Let $t\mapsto f_t \in\Psp_1\cap\Csp^1$, 
be a classical solution of \rVPa\ over some time interval $t\in [0,T)$,
and assume that	initially and hence for all $t\in [0,T)$ we have
$\norm{f_t}_{3/2}<\frac{3}{8} \left(\frac{15}{16}\right)^{1/3}$.
	Then, uniformly in $t$, we have
\bea
 \norm{\pdq\phi_t}_2^2 
&\leq &
8\pi\varkappa(f_0)  \cE(f_0) 
\,,
\label{coroBOUNDSfOFtEpot}
\\
\iint \sqrt{1+|p|^2} f_t(p,q)\dd{p}\dd{q}
&\leq&
 (1+\varkappa(f_0)) \cE(f_0) 
\,,
\label{coroBOUNDSfOFtEkin}
\eea
and, for all $\alpha\in[1,\infty]$, 
\beq
\hskip 3truecm
\norm{\rho_t}_{\gamma} 
\leq 
C(\alpha) \|f_0\|_\alpha^{\eta} (1+\varkappa(f_0))^{1-\eta}\cE(f_0)^{1-\eta}
\label{rhoOFtSUBcritLbetaBOUND}
\eeq
with $\gamma$ and $\eta$ given by \refeq{betaANDeta}.
\end{Coro}
\noindent\textit{Proof of Corollary \ref{coroBOUNDSfOFt}.}
	Under the hypotheses of Corollary \ref{coroBOUNDSfOFt}, which imply
the conservation of energy, $\cE(f_t) = \cE(f_0)$, and of the $\Lsp^\alpha$ norms,
$\norm{f_t}_\alpha =\norm{f_0}_\alpha$ (note that any function $f\in\Psp_1\cap\Csp^1$ 
is automatically in all $\Lsp^\alpha$), the uniform bounds 
\refeq{coroBOUNDSfOFtEpot}--\refeq{coroBOUNDSfOFtEkin} follow at once from Corollary
\ref{coroBOUNDSf}, and \refeq{rhoOFtSUBcritLbetaBOUND} from Corollary \ref{coroRHOsubCRITbound}.
\QED
\begin{Theo}\label{theoEXISTS}
	Let $t\mapsto f_t\in\Psp_1\cap\Csp^1$ be a classical solution of \rVPa\ over 
some interval $t\in [0,T)$, launched by spherically symmetric data $f_0$ with compact support 
in $p$-space, vanishing for $p\times{q}=0$, and assume that initially and hence for all $t\in [0,T)$ 
we have $\norm{f_t}_{3/2}<\frac{3}{8} \left(\frac{15}{16}\right)^{1/3}$.
       Then the momentum support is uniformly bounded on $[0,T)$ and hence \rVPa\
possesses a global classical solution.
\end{Theo}

	The proof of the above, which mimics closely the argument for global existence given in 
\cite{GlasseySchaefferA}, hinges on the following estimate:

\begin{Lemm}\label{lemmPhiEst} 
	Let $t\mapsto f_t$ be a classical solution of \rVPa over some interval $t\in [0,T)$.
	Then for any pair of exponents $\gamma<3<\alpha$ there exists a constant $C_{\alpha,\gamma}$ such that
\beq
|\pdq \phi_t| \leq C_{\alpha,\gamma} \|f_0\|_\alpha^\theta \|\rho_t\|_\gamma^{1-\theta} P^\xi(t)
\eeq
where 
\beq 
P(t) :=  \sup\left\{|p|\big| (p,q) \in \supp(f_s),\ 0\leq s\leq t\right\} 
\eeq
and 
\beq
\theta := \frac{1 - \gamma/3}{1 - \gamma/\alpha} \in (0,1),
\qquad
\xi := 3(1-{1}/{\alpha})\theta
\eeq
\end{Lemm}

\noindent
\textit{Proof of Lemma \ref{lemmPhiEst}.}
	We have that $\phi_t = -|\id|^{-1} * \rho_t$ and hence
\begin{eqnarray*}
|\pdq \phi_t(q)|
& \leq & \int_{|q-q^\prime|<R} \frac{\rho_t(q^\prime)}{|q-q^\prime|^2} \dd{q^\prime} +
 \int_{|q-q^\prime|\geq R} \frac{\rho_t(q^\prime)}{|q-q^\prime|^2} \dd{q^\prime} \\
& \leq &  c_1^\prime(\alpha) \|\rho_t\|_\alpha R^{1-3/\alpha} + c_2(\gamma) \|\rho_t\|_\gamma R^{1-3/\gamma} \\
& \leq & c_1(\alpha) \|f_t\|_\alpha P(t)^{3-3/\alpha}R^{1-3/\alpha} + c_2(\gamma) \|\rho_t\|_\gamma R^{1-3/\gamma},
\end{eqnarray*}
which upon optimizing in $R$ gives the desired result. 
\QED

\noindent
\textit{Proof of Theorem \ref{theoEXISTS}:}
	By spherical symmetry, $f_t (p,q) = \overline{f}_t(|p|,|q|,p\cdot q)$ and thus
$\rho_t(q) = \overline{\rho}_t(|q|)$ so that we have $\pdq \phi_t(q) = M(|q|,t) |q|^{-3} {q}$, where
\beq
M(|q|,t) : = 4\pi \int_0^{|q|} \overline{\rho}_t(r)r^2  \dd{r}.
\eeq
	 Note that $\lim_{|q|\to \infty} M(|q|,t) = 1$.
	Let the exponent $\alpha>3$ be fixed, and let $\gamma  = \frac{4\alpha -3}{3\alpha-2}$.
	Thus $9/7<\gamma\leq 4/3$.
	From Lemma \ref{lemmPhiEst} and Corollary \ref{coroBOUNDSfOFt} we then have that in the 
spherically symmetric case
\beq
|\pdq\phi_t(q)|
\leq \min\{|q|^{-2} , C P^\xi(t) \} \leq 4(C^{-1/2} (P(t))^{-\xi/2} +|q|)^{-2}
\eeq
where the constant $C$ depends only on the initial data $f_0$ (specifically on its energy,  its
$\Lsp^{3/2}$ norm, and its $\Lsp^\alpha$ norm).

	The next step, following the Glassey-Schaeffer argument, is to analyze the characteristics 
of the Vlasov equation in the spherically symmetric case, to conclude that, for any $T_0 \in [0,T)$,
\beq
 P(T_0) \leq P(0) + \sqrt{1+P^2(0)} + C^{1/2} P(T_0)^{\xi/2}
\,.
\eeq
	We now compute:
\beq 
 \xi 
  = 
   3\left(1-\frac{1}{\alpha}\right)
	\frac{1 - {\gamma}/{3}}{1-{\gamma}/{\alpha}} 
  = 
 \frac{5\alpha - 3}{3 \alpha - 3}
\,.
\eeq
	Since $\xi < 2$ for $\alpha>3$, this implies a uniform bound on $P(T_0)$ regardless of the 
size of $C$, and since $T_0$ was arbitrary in $[0,T)$, the theorem follows.
\QED
\section{Supercritical solutions}
	Recall that supercritical solutions are those for which 
$\norm{f_t}_{3/2} > \frac{3}{8} \left(\frac{15}{16}\right)^{1/3}$.
	Notice that the energy of a supercritical solution can actually be strictly positive, zero, 
or strictly negative, and arbitrarily negative at that.
        While not much seems to be known about supercritical
solutions with positive energy, \emph{if} the energy of a solution is non-positive, i.e. if $\cE(f_t)\leq 0$
(which implies that the solution is necessarily supercritical by Proposition \ref{propEofFbound}),
then one can rule out certain classes of global solutions.

        In particular, when $\cE(f_0)\leq 0$ we can rule out that $f_t$ is stationary.
	This is an immediate consequence of the stationary virial theorem, 
Corollary \ref{coroVIRIALthm}, which asserts that $\cE(f)>0$ for any stationary solution.
	Since stationary solutions are global, we thus have ruled out a whole subclass of 
global solutions when $\cE(f_0) \leq 0$. 
	Note that the arguments just presented work without any symmetry assumption on $f$.

	The dynamical virial theorem can be used to rule out further types of global solutions $f_t$ 
with $\cE(f_0) \leq 0$.
        By Theorem III of Glassey and Schaeffer \cite{GlasseySchaefferA}, for $\cE(f_0)<0$ no spherical 
solution with compact support is global.
	We now prove the following generalization of the blow-up result in
\cite{GlasseySchaefferA}, which states that also some data with $\cE(f_0)=0$
lead to finite time blow-up.
\begin{Theo}\label{theoBLOWUP} 
	Let $t\mapsto f_t\in\Psp_3\cap\Csp^1$ be a spherically symmetric supercritical classical
solution of \rVPa\ over some interval $[0,T)$, and assume that $\cE(f_0) \leq 0$.
	In case that $\cE(f_0)= 0$, also assume that $\cV(f_0)\leq -1/2$.
	Then $T<\infty$. 
\end{Theo}
\noindent
\textit{Proof of Theorem \ref{theoBLOWUP}.}
	For $\cE(f_0) < 0$ this was proved (for compactly supported $f_0$) in \cite{GlasseySchaefferA}.
	Inspection of their proof of their Theorem~III shows that it applies verbatim here, where
compactness is replaced by the $\Psp_3$ assumption.

	In case that $\cE(f_0)= 0$ we argue as follows. 
	By \refeq{dilationID} we find that $\cV(f_t)<\cV(f_0)$ for all $t\in[0,T)$,
and since by hypothesis $\cV(f_0)\leq -1/2$, we conclude that
$\cV(f_t)< -1/2$ for all $t\in[0,T)$; in fact, even 
$\cV(f_t)< -C < -1/2$ for all $t\in(\eps,T)$ (Note that $C$ will depend on $\eps$ and on $f_0$, but
this is irrelevant for the argument).
	Next, as estimated in \cite{GlasseySchaefferA} in their proof of Thm.~III, sphericity implies 
that $\abs{\iint  |q|^2 v\cdot\pdq\phi_t(q)f_t(p,q) \dd{p}\,\dd{q}}\leq 1$.
	Thus, by integrating \refeq{virialIDtwo} over $t$ from $0$ to $T$, and then using the
above estimates, we find that 
\beq
\!\!\!\iint\! |q|^2 \sqrt{1+ |p|^2} f_T(p,q) \dd{p}\dd{q}
\leq 
\!\!
\iint\! |q|^2 \sqrt{1+ |p|^2} f_0(p,q) \dd{p}\dd{q} + 2C\eps - (2C-1)T 
,
\label{virialAspinoff}
\eeq
and since $2C-1 >0$, it follows that the r.h.s. of \refeq{virialAspinoff} $<0$
when $T$ is large enough, while the l.h.s. is strictly positive. 
	Therefore, $T$ cannot be too large.\QED
\begin{Rema}
        We don't know whether the condition $\cV(f_0)\leq -1/2$ is actually sharp.
\end{Rema}
\begin{Rema}
        Classical solutions of nonrelativistic VP$^-$ don't blow up \cite{Pfaff,Schaeffer}.
\end{Rema}
\section{Critical solutions}
	Our global existence proof of classical solutions to the Cauchy problem for subcritical
data relies heavily on the a-priori estimates for subcritical solutions obtained
from a-priori bounds on the individual energy functionals $\cE_p(f)$ and $\cE_q(f)$.
        When $f$ is critical so that
$\norm{f_t}_{3/2} = \frac{3}{8} \left(\frac{15}{16}\right)^{1/3}$, 
then our Proposition \ref{propEofFbound} still guarantees that
$\inf\cE(f_t)=0$ and further that $\inf\cE(f)\neq\min\cE(f)$ 
over the set of critical data, but these estimates do not seem to produce
a-priori bounds on the $p$-space and $q$-space energy functionals $\cE_p(f)$ and $\cE_q(f)$.
        So no global existence and uniqueness proof for critical classical solutions
seems in sight.
        Curiously, no finite-time blow-up result is available either.
\section{Some further interesting open problems}
	In addition to the wide open questions about critical solutions, there are a number of
interesting problems that we would like to draw attention to.

        Some of the assumptions in our global existence and uniqueness Theorem \ref{mainTHEOREM}
are adapted from \cite{GlasseySchaefferA}, in particular the compact $p$-support condition 
and the condition that $f_0$ vanish for $p\times{q}=0$. 
        We expect that the compact support condition can be replaced by a weaker one, 
like sufficiently rapid decay ``at infinity,'' and that $f_0=0$ for $p\times{q}=0$ can be dropped.

        Since all solutions for which global existence and uniqueness has been proved are subcritical,
one may want to consider the reverse question, whether a unique spherically symmetric global 
classical solution necessarily has to be subcritical. 
	We suspect that the answer to this reverse problem is negative, but the problem is
open. 
	If the answer to this reverse problem is negative, it may be forthcoming more readily 
by studying the special subset of global-in-time solutions furnished by the stationary solutions 
to \rVPa, of which there are many examples (see \cite{Batt, HadzicRein}). 

	Then there is the open question of the sharp values of $C_\beta$ when $\beta>3/2$.

	Another interesting question is that of the weakest possible $\Lsp^\alpha$ norm for 
classical data which controls $\nabla\phi$.
	Our global existence and uniqueness proof for Theorem \ref{mainTHEOREM} 
invokes $f_0\in\Lsp^\alpha$ for any $\alpha>3$, beside subcriticality.
	We suspect that $\alpha=3$ may be the critical $\alpha$ value for this question, indeed,
but it would be good to have a definitive answer.
        A variant of this question should become particularly relevant when one asks for weaker 
solutions than classical, as e.g. in \cite{DiPernaLions}.

        Finally, as pointed out earlier, our non-existence result for stationary solutions with 
$\cE(f)\leq 0$ does not assume spherical symmetry.
        So one wonders how much of our other results generalizes to non-spherical solutions.
	This may seem primarily of mathematical interest, for without the
sphericity assumption we are in danger of leaving the realm of physical validity of \rVPa. 
	Yet, since spherical symmetry is never a perfect symmetry of nature, it is
important to show that significant qualitative results do not sensitively depend on having exact 
spherical symmetry. 
\medskip

\textbf{Note added}
	After our paper was accepted for publication we discovered that our open question of the sharp 
values of $C_\beta$ for $\beta>3/2$ is implicitly answered by Proposition 1.1 of \cite{LMRa} which
characterizes a subset of the compact Lane-Emden polytropes  as optimizers in the variational principle (1.14) 
of \cite{LMRa}, to which our variational principle \refeq{CbetaDEF} in Proposition 
\ref{propEofFbound}, when restricted to $\beta>3/2$, is equivalent.
	The quantitative evaluation of $C_\beta$, which requires numerical integration of the Lane-Emden 
polytrope equation, was meanwhile carried out by Brent Young and will be reported on elsewhere.
\medskip

\textbf{Acknowledgement} 
Kiessling was supported by NSF Grants DMS-0103808 and DMS-0406951, 
and in parts by CNRS through a poste rouge to him while visiting CNRS-Universit\'e de Provence;
Tahvildar-Zadeh was supported by NSF Grant DMS-0301207. M.K. thanks Yves Elskens and Elliott Lieb 
for their comments on the proofs of Props. \ref{propEboundOPT} and \ref{propEofFbound}. Thanks
go also to Rupert Frank and an anonymous referee for drawing our attention to \cite{LMRb}, 
which in turn led us to \cite{LMRa}.
%
%
\section*{APPENDIX}
{\textbf{\ \ \ \ \ \ \ \ Derivation of \rVPa\ with spherical symmetry from rVM}}

\noindent
	We begin by recalling the relativistic Vlasov--Maxwell equations for a two species 
plasma, containing a specie  of $N^+$ positively, and another one of $N^-$ negatively  
charged particles.
	For simplicity, it is assumed that all particles carry the same \emph{magnitude}
of charge and the same mass, as in an electron-positron plasma, and that there is overall
an even number of particles so that $N^-+N^+=2N$.
	Units are chosen such that mass and magnitude of charge both $=1$.
	Then, the particle density functions $f^\pm_t(p,q)\in\Psp_1\cap\Csp^1$ satisfy
\begin{equation}
\pdt f_t^\pm(p,q)
+ 
v \cdot\pdq f_t^\pm(p,q)\pm\Big(E_t(q) + v\times B_t(q)\Big)\cdot\pdp f_t^\pm(p,q)
= 0\,,
\label{eq:rVMfEQs}
\end{equation} 
where velocity $v$ and momentum $p$ are related by Einstein's formula \refeq{EINSTEINvOFp}, while
the electric field $E_t(q)\in(\Lsp^2\cap\Csp^1)(\dd{q})$ and the magnetic field 
$B_t(q)\in(\Lsp^2\cap\Csp^1)(\dd{q})$ at the space 
point $q\in{\Rset}^3$ at time $t\in{\Rset}$ satisfy the evolution equations 
\begin{eqnarray}
        \pdt{{B}_t(q)}
&= &
        - \pdq\times{E}_t(q)   
\, ,
\label{eq:rVMrotE}
\\
         \pdt{{E}_t(q)}
&= &
\pdq\times{B}_t(q)    - 4\pi  \int_{{\Rset}^3}v (\nu^+f_t^+ -\nu^-f_t^-)(p,q)\dd{p}\,,
\label{eq:rVMrotB}
\eea
supplemented by the constraint equations
\bea
        \pdq\cdot {B}_t(q)  
&= &
        0\, ,
\label{eq:rVMdivB}
\\
        \pdq\cdot{E}_t(q)  
&=&
        4 \pi \int_{{\Rset}^3} (\nu^+f_t^+ -\nu^-f_t^-)(p,q)\dd{p}\,.
\label{eq:rVMdivE}
\end{eqnarray}
	Here, $\nu^\pm$ are the relative numbers of charges in the positive and negative specie; 
i.e.,\footnote{Note that in \refeq{eq:rVMfEQs}--\refeq{eq:rVMdivE} we have chosen 
		time and space scaled with $2N$ rather than $N$ as in rVP.}
$N^-=\nu^- 2N$ and $N^+=\nu^+ 2N$.
	Global well-posedness of the Cauchy problem for small initial data is known, and for 
large data under the additional assumption that no singularities occur near the light cone,
see~\cite{GlasseySchaefferB,GlasseyStraussA,GlasseyStraussB,KlaSta} and many further references 
therein.
	Note that the constraints propagate when satisfied by the initial data $E_0(q)$ and $B_0(q)$.

	If we set \emph{either} $\nu^+=0$ and $\nu^-=1$ \emph{or} $\nu^-=0$ and $\nu^+=1$,
and ignore the pertinent $f^+$, respectively $f^-$ equation from the pair \refeq{eq:rVMfEQs}, 
the equations \refeq{eq:rVMfEQs}--\refeq{eq:rVMdivE} of two-species rVM reduce to one-specie rVM. 
	All spherically symmetric solutions of repulsive rVP (i.e. \rVPr, viz.
$\sigma=+1$ in \refeq{rVPfEQ}--\refeq{phiASqTOoo}) also solve this 
one-specie rVM.
	Indeed,  one-specie rVM simply reduces to \rVPr\ in the special situation of spherical 
symmetry, see \cite{HorstB}.
	The reasons are: (i) all magnetic effects, in particular all electromagnetic waves,
vanish in a spherically symmetric solution of rVM, rendering the charged particle interactions 
purely electric; and (ii) any two electric charges of the same sign repel each other electrically.

	By contrast, spherically symmetric solutions of attractive rVP (i.e. \rVPa, viz.
$\sigma=-1$ in \refeq{rVPfEQ}--\refeq{phiASqTOoo}) do \emph{not} solve rVM for any choice of
$\nu^+$ and $\nu^-$, nor for any other number of electrically charged particle species, so that 
any attempt to derive \rVPa\ from rVM for two charged species of electrical particles would 
seem entirely misguided. 
	However, we will now argue (convincingly, we hope) that certain families of 
\emph{distributional} solutions of overall neutral two-species rVM (without spherical 
symmetry) converge to spherically symmetric solutions of \rVPa. 
	Here is our argument.

	Consider now an overall neutral two-species plasma, so $N^+=N^- = N$, i.e.
$\nu^+ = \nu^- = 1/2$.
	We are interested in the distributional solutions representing the \emph{actual}
empirical ``densities'' of the underlying $2N$-body system, which are sums of atomic measures
(given below).
	To be able to work with such singular measures, we first need to regularize the equations
of rVM by convolution with a smooth positive density function $\fe$ with $SO(3)$ symmetry and 
compact support, satisfying $\int_{{\Rset}^3}\fe({q}){\rm d}^3q = 1$, which will be removed at
the end of our construction.\footnote{Note that this regularization breaks the Lorentz covariance; 
		however, we do not invoke any Lorentz transformations in our reasoning, and for 
		continuum solutions, when we let $\fe\to\delta$, we formally recover the Lorentz 
		covariant relativistic Vlasov--Maxwell equations.}
	The regularized rVM reads\footnote{The unfamiliar factors $2\pi$ 
		in \refeq{regVMrotB} and \refeq{regVMdivE} 
are a consequence of $\nu^-= 1/2=\nu^+$, 
		which have been factored out from under the integrals and 
		multiplied into $4\pi$.}
\begin{equation}
\pdt \mu_t^\pm(p,q)
+ 
v \cdot\pdq \mu_t^\pm(p,q)
\pm\Big((\fe*E_t)(q) + v\times (\fe*B_t)(q)\Big)
\cdot\pdp \mu_t^\pm(p,q)
= 0\,,
\label{regVMfEQs}
\end{equation} 
\begin{eqnarray}
        \pdt{B_t(q)}
        + \pdq\times E_t(q)   
&= &
        0\, ,
\label{regVMrotE}
\\
        - \pdt{E_t(q)}
        + \pdq\times B_t(q)  
&= &
        2\pi \int_{{\Rset}^3}v (\fe*(\mu_t^+ -\mu_t^-))(p,q)\dd{p}\,,
\label{regVMrotB}
\\
        \pdq\cdot B_t(q)  
&= &
        0\, ,
\label{regVMdivB}
\\
        \pdq\cdot E_t(q)  
&=&
        2 \pi \int_{{\Rset}^3} (\fe*(\mu_t^+ -\mu_t^-))(p,q)\dd{p}
\,.
\label{regVMdivE}
\end{eqnarray}
	Here we introduced the notation $\mu_t^\pm$ which could either mean density functions
$f^\pm_t\in\Psp_1\cap\Lsp^1$ as before, or true measures $\mu_t^\pm\in \Psp_1$.
	In particular, we can allow $\mu^\pm_0$ to be the \emph{empirical relative ``densities''}
on $\Rset^6$ at time $t=0$ of the underlying $2N$-body system, defined as follows.
	Letting particles with even index be positively charged and those with 
odd index negatively, we can identify any point $(p_{2k},q_{2k})_{k=1}^N\in\Rset^{6N}_{even}$ 
($=N$-positive-charges subspace of $2N$-body phase space) with a singular 
empirical relative ``density'' on $\Rset^6$, 
\begin{equation}
\triangle^+(p,q)
 = \frac{1}{N} \sum_{k=1}^N \delta(p-p_{2k})\delta(q-q_{2k}),
\label{empDENSpos}
\end{equation}
and each point $(p_{2k-1},q_{2k-1})_{k=1}^N\in\Rset^{6N}_{odd}$ with a singular empirical 
relative ``density''
\begin{equation}
\triangle^-(p,q)
 = \frac{1}{N} \sum_{k=1}^N \delta(p-p_{2k-1})\delta(q-q_{2k-1})
\label{empDENSneg}
\,.
\end{equation}
	We write $(p_{k}(0),q_{k}(0))_{k=1}^{2N}\in\Rset^{12N}$ if the point in $2N$-body phase
space is the initial phase point (at time $t=0$) of the phase space trajectory
$t\mapsto (p_{k}(t),q_{k}(t))_{k=1}^{2N}$ of our plasma, and the corresponding empirical
densities (dropping the quotes from now on) are denoted $\triangle^\pm_0(p,q)$, respectively
$\triangle^\pm_t(p,q)$.
	Let the $\mu^\pm_0$ be given by some $\triangle^\pm_0(p,q)$. 
	Then these initial empirical densities together with compatible initial data for
the fields, $E_0(q)$ and $B_0(q)$, launch a subsequent evolution under 
\refeq{regVMfEQs}--\refeq{regVMdivE} for which the $\mu^\pm_t$ are also given by empirical 
densities, viz. $\mu^\pm_t(p,q) = \triangle^\pm_t(p,q)$, characterized as follows.

	Let the evolution equations for the dynamical variables of each particle, i.e. 
position $q_k(t)$ and linear momentum $p_k(t)$ be given by
\begin{equation}
 \frac{\dd{q}_k}{\dd{t}}\Big|_{q_k=q_k(t)}
  = 
 \frac{p_k(t)}{\sqrt{1+|p_k(t)|^2}}
\,,
\label{eq:pDEF}
\end{equation} 
\begin{equation}
        \frac{\dd{p_k}}{\dd{t}}\Big|_{p_k=p_k(t)}
= 
        e_k\left[(\fe*E_t)(q_k(t)) + \dot{q}_k(t) \times(\fe*B_t)(q_k(t))\right]
\,,
\label{eq:ALeq}
\end{equation}
with $e_k=-1$ if $k$ is odd, and $e_k=+1$ if $k$ is even.
	The above equations are the Einstein--Newton equations of motion, 
equipped with the Abraham--Lorentz expressions for the volume-averaged Lorentz force that
acts on each particle.
	The evolution equations for the electric field $E_t(q)$ and 
the magnetic field $B_t(q)$ now read
\bea
        \pdt{B_t(q)}
        + \pdq\times{E}_t(q)   
&= &
        0\, ,
\label{eq:MLrotE}
\\
        - \pdt{E_t(q)}
        + \pdq\times{B}_t(q)  
&= &
        2\pi \frac{1}{N}\sum_{k=1}^{2N} e_k\fe({q}-{q}_k(t))\, \dot{q}_k(t) 
\,,
\label{eq:MLrotB}
\eea
and the constraint equations are
\bea
        \pdq\cdot {B}({q}, t)  
&= &
        0\, ,
\label{eq:MLdivB}
\\
        \pdq\cdot{E}({q}, t)  
&=&
       2\pi  \frac{1}{N}\sum_{k=1}^{2N} e_k\fe({q}-{q}_k(t))
\,,
\label{eq:MLdivE}
\eea
altogether known as the classical Maxwell--Lorentz field equations; note that
\refeq{eq:MLrotB} and \refeq{eq:MLdivE} are just \refeq{regVMrotB} and \refeq{regVMdivE} 
with $\mu^\pm_t$ given by $\triangle^\pm_t$. 
	It was proved recently in
\cite{BauerDuerr, KomechSpohn} that the dynamical equations  \refeq{eq:pDEF}--\refeq{eq:MLdivE}
are globally well posed as Cauchy problem in convenient Hilbert spaces; see also 
\cite{KunzeSpohnC, SpohnBOOK}.
	Let $t\mapsto (p_{k}(t),q_{k}(t))_{k=1}^{2N}\in\Rset^{12N}$ be the particle phase space 
trajectory of a global finite energy solution to this Abraham--Lorentz\footnote{These 
		semi-relativistic equations  of Abraham--Lorentz electrodynamics are a non-Lorentz 
		covariant regularization of formally Lorentz covariant formal Lorentz electrodynamics 
		with point charges (which has the unpleasant feature that its formal equations are 
		mathematically ill-defined without regularization). 
		The semi-relativistic Abraham--Lorentz model is chosen purely for the ease of the 
		discussion.
		A fully Lorentz covariant regularized Lorentz model is available 
		(see \cite{AppKieAOP, AppKieLMP, SpohnBOOK}) but is considerably more complicated.}
model \refeq{eq:pDEF}--\refeq{eq:MLdivE}.
	Then the corresponding dynamical empirical densities $\triangle^\pm_t(p,q)$ satisfy 
regularized rVM \refeq{regVMfEQs}--\refeq{regVMdivE} in the sense of distributions.
	Thus the Abraham--Lorentz model is entirely equivalent to the regularized rVM 
\emph{restricted to empirical densities} $\triangle^\pm_t$.
	
	We now note a very important point about the relationship of distributional solutions
of (regularized) rVM and its continuum solutions.
	Recall that in the introduction we pointed out that $f_t$ ``should really be thought of 
as a continuum approximation to a \emph{merely normalized} (i.e. relative) empirical ``density'' on 
$(p,q)$-space of an actual individual $N$-body system.''
	In this sense, assume that $N$ is sufficiently large so that for both species
$\triangle^\pm_0 \approx f_0^\pm\in\Psp_1\cap \Lsp^1$ closely in measure (i.e. w.r.t. some 
Kantorovich-Rubinstein distance).
	Then \emph{on a suitably short time scale} the evolutions of the $\triangle^\pm_t$ under
\refeq{regVMfEQs}--\refeq{regVMdivE} will be reasonably closely approximated by solutions of 
\refeq{regVMfEQs}--\refeq{regVMdivE} with the initial data $\triangle^\pm_0$ replaced by 
$f^\pm_0\in \Psp_1\cap\Lsp^1$, and with the initial data for the fields, $E_0$ and $B_0$,  
replaced accordingly.\footnote{For a rigorous proof of this for a scalar caricature of rVM, see \cite{EKR}.}
	On longer time scales, various deviations of the rVM evolution for regular initial data
will become visible, and as we shall see now, under favorable conditions, one of those 
long-time evolutions is captured precisely by regularized \rVPa.
	For continuum solutions we can subsequently let $\fe\to\delta$, recovering \rVPa.

	To see this, contemplate that both $\triangle^\pm_0 \approx f_0^\pm\equiv f_0$, with $f_0$
spherical.
	Then on the conventional short Vlasov time scale we will just find non-interacting perfect gas
dynamics, for $f^+_t - f^-_t \equiv 0$, then, in \refeq{eq:rVMrotB} and \refeq{eq:rVMdivE} 
with $\nu^-=1/2=\nu^+$.
	However, if \emph{all} the particles are i.i.d. by $f_0$, then on a longer time 
scale we should obtain \rVPa\ for $f_t$, and here is why.
	First of all, by (approximate) spherical symmetry of the $2N$-body plus field system
we expect that magnetic effects can again be neglected, so we set $B\equiv 0$.
	But then \refeq{eq:MLdivE}, \refeq{eq:MLrotE},  and \refeq{eq:MLrotB} are solved 
by Coulomb's formula
\beq
E_t(q) =
-
 \frac{1}{2N}\sum_{k=1}^{2N}e_k \pdq \left(|\id|^{-1}*\fe(\,\cdot\,-{q}_{k}(t))\right)(q)
\eeq
	Evaluating r.h.s. \refeq{eq:ALeq} with this formula for $E$, setting $B\equiv 0$,
we find for particle $\ell$,
\bea
        e_\ell(\fe*E_t)(q_\ell(t)) =
\!\!\!&-&\!\!\!
 \frac{e_\ell}{2N}\sum_{k=1}^{N}\left(\fe*
 \pdq \left(|\id|^{-1}*\fe(\,\cdot\,-{q}_{2k}(t))\right)\right)(q_\ell(t))
\nonumber
\\
\!\!\!&+&\!\!\!
 \frac{e_\ell}{2N}\sum_{k=1}^{N}\left(\fe*
\pdq \left(|\id|^{-1}*\fe(\,\cdot\,-{q}_{2k-1}(t))\right)\right)(q_\ell(t))
\label{regCOULforce}
\eea
where we made use of our convention that positive particles carry an even, negative an odd index.
	Now observe that since by hypothesis \emph{all} particles are i.i.d. by $f_0$, all \emph{but one}
of the force terms in \refeq{regCOULforce} are i.i.d. random variables, the exception being 
\beq
\left(\fe*
\pdq \left(|\id|^{-1}*\fe(\,\cdot\,-{q}_\ell(t))\right)\right)(q_\ell(t)) \equiv 0,
\eeq
which states that the Coulomb self-force on the particle vanishes. 
	But then, if $\ell$ is even, a term from the first sum is missing, and if $\ell$ is odd,
a term from the second sum is missing.
        For all the other terms, since \emph{all} particles are i.i.d., we can use that both 
$\triangle^\pm_t\approx f_t$, and paying attention to the correct normalization,\footnote{Mathematically 
		this is a nice instance where one might be tempted to set two extremely huge numbers $N$ 
                and $N-1$ equal, but here their difference matters, and $N-(N-1) =1\neq 0 = N-N$.}
we now find that
\beq
e_\ell(\fe*E_t)(q_\ell(t)) 
 \approx
	 \frac{1}{2N}\left(\fe* \pdq \left(|\id|^{-1}*(\fe* \rho_t)\right)\right)(q_\ell(t))
\eeq
which is independent of $e_\ell$: this means that each particle is acted on by a net attractive central 
force, for from each particle's perspective the rest of the system is singly oppositely charged, as the 
rest of the system always contains one more of the oppositely charged than the equally charged particles; 
the force is (approximately) central by (approximate) spherical symmetry. 
	We therefore introduce
\beq
\phi_t^\eps 
: =
	- \frac{1}{2N} \fe* \left(|\id|^{-1}*(\fe* \rho_t)\right)
\label{regEFFphi}
\eeq
replace $e_\ell E_t(q_\ell(t))$ by $-\pdq\phi_t^\eps(q_\ell(t))$ for both positive and
negative charges in the regularized rVM (noting that $\pdq$ and $\fe*$ commute) and 
also replace $\triangle^\pm_t$ by $f_t$, upon which both equations \refeq{regVMfEQs} reduce to the same 
regularized Vlasov equation
\beq
 \Big(\pdt + v \cdot\pdq -\pdq\phi_t^\eps(q)\cdot\pdp\Big)f_t(p,q) = 0,
\label{rVPfEQappREG}
\eeq
in which the velocity $v\in\Rset^3$ and momentum $p\in\Rset^3$ of a (point) particle of unit mass
are again related by Einstein's formula \refeq{EINSTEINvOFp}, and where
$\phi_t^\eps$ is given by the r.h.s. of \refeq{regEFFphi}.
        Note that \refeq{rVPfEQappREG} is decoupled from the Maxwell-Lorentz field equations.
	If $f_t$ is sufficiently regular, and uniformly so for all $\eps$, 
we may now let $\fe\to\delta$ and find that the resulting $f_t$ is a solution to the Vlasov equation
\beq
 \Big(\pdt + v \cdot\pdq -\pdq\phi_t(q)\cdot\pdp\Big)f_t(p,q) = 0,
\label{rVPfEQapp}
\eeq
where now
\beq
\phi_t(q) = - \frac{1}{2N} 
\left(|\id|^{-1}* \rho_t\right)(q).
\eeq
	Clearly, the scalar field $\phi_t$ satisfies the Poisson equation
\beq
 \pdqsq\phi_t(q) =4\pi \frac{1}{2N} \int_{\Rset^3}\! f_t(p,q)\dd{p}
\label{rVPphiEQapp}
\eeq
with asymptotic condition
\beq
 \phi_t(q) \asymp - (2N |q|)^{-1}
\,
\label{phiASqTOooAPP}
\eeq
when $|q|\to\infty$. 
	By a final rescaling of space and time variables we can get rid of the factor $1/2N$
and thus have obtained \rVPa.
	
\begin{Rema} 
	The i.i.d. assumption on \emph{all} particles is very important. 
	Indeed, even if we merely assume that the particles of each specie are i.i.d. w.r.t. 
$f_0$ separately, then we can still have that particles of opposite species are strictly 
correlated, viz. wherever a positive particle is located, a negative one is, too. 
	This is the classical analog of a neutral gas of ``positronium atoms,''
and no electric \rVPa\ will result.
\end{Rema}
\newpage
%

%
\end{document}